\documentclass[%
superscriptaddress,
reprint,
showpacs,
amsmath,amssymb,
aps,
pre,
]{revtex4-1}

\usepackage{graphicx}
\usepackage{dcolumn}
\usepackage{bm}
\usepackage{color,verbatim}

\begin {document}
\title{Ergodic properties of continuous-time random walks: finite-size
  effects and ensemble dependences}

\author{Tomoshige Miyaguchi}
\email{tmiyaguchi@naruto-u.ac.jp}
\affiliation{%
  Department of Mathematics Education, Naruto University of Education, Tokushima 772-8502, Japan
}
\author{Takuma Akimoto}
\email{akimoto@z8.keio.jp}
\affiliation{%
  Department of Mechanical Engineering, Keio University, Yokohama 223-8522, Japan
}%

\date{\today}


\begin{abstract}
  The effects of spatial confinements and smooth cutoffs of the waiting
  time distribution in continuous-time random walks (CTRWs) are studied
  analytically. We also investigate dependences of ergodic properties on
  initial ensembles (i.e., distributions of the first waiting time). Here,
  we consider two ensembles: the equilibrium and a typical non-equilibrium
  ensembles. For both ensembles, it is shown that the time-averaged mean
  square displacement (TAMSD) exhibits a crossover from normal to anomalous
  diffusion due to the spacial confinement and this crossover does not
  vanish even in the long measurement time limit. Moreover, for the
  non-equilibrium ensemble, we show that the probability density function
  of the diffusion constant of TAMSD follows the transient Mittag-Leffler
  distribution, and that scatter in the TAMSD shows a clear transition from
  weak ergodicity breaking (an irreproducible regime) to ordinary ergodic
  behavior (a reproducible regime) as the measurement time increases. This
  convergence to ordinary ergodicity requires a long measurement time
  compared to common distributions such as the exponential distribution; in
  other words, the weak ergodicity breaking persists for a long time. In
  addition, it is shown that, besides the TAMSD, a class of observables
  also exhibits this slow convergence to ergodicity. We also point out
  that, even though the system with the equilibrium initial ensemble shows
  no aging, its behavior is quite similar to that for the non-equilibrium
  ensemble.
\end{abstract}

\maketitle


\section {Introduction}

Recently, slow anomalous diffusion, which is defined by the sublinear
dependence of the mean square displacement (MSD) on time, has been found in
various phenomena including lipid granules diffusing in living fission
yeast cells \cite{jeon11, burov11}, colloidal particles diffusing on sticky
surfaces \cite{xu11} and in networks of entangled actin filaments
\cite{wong04}, mRNA molecules diffusing in {\it E. coli} \cite{golding06},
chromosomal loci diffusing in bacteria \cite{weber10, *weber10a}, telomeres
diffusing in nuclei of eukaryote cells \cite{bronstein09, *kepten11}, and
proteins diffusing in dextran solutions \cite{banks05}. To understand these
phenomena, two types of slow diffusion models have been extensively studied
so far: (1) continuous-time random walks (CTRWs) \cite{jeon11, burov11,
  xu11, wong04} and (2) the generalized Langevin equation (GLE) and
fractional Brownian motions (FBMs) \cite{golding06, magdziarz09,
  *magdziarz10, weber10, *weber10a, bronstein09, *kepten11, banks05}. For
the CTRW model, if the probability density function (PDF) of the waiting
times between jumps of the particle is a power law $w(\tau) \sim
1/\tau^{1+\alpha}$ with $0<\alpha<1$, the ensemble-averaged MSD (EAMSD)
shows slow diffusion \cite{scher75, bouchaud90}: $\left\langle (\delta
x)^{2}\right\rangle \sim t^{\alpha}$%
\footnote{%
  In this work, we use ``$\sim$'' to stand for an asymptotic relation,
  i.e., $f(s) \sim g (s)$ means $f(s) / g(s) = \mathcal{O}(1)$ for some
  asymptotic limit such as $s \to 0$. On the other hand, we use
  ``$\simeq$'' to stand for a stricter asymptotic relation: $f(s) \simeq g
  (s)$ means $f(s) / g(s) = 1$ for some asymptotic limit. Furthermore, we
  use ``$A \approx B$'' when $A$ is an approximation of $B$, and ``$A
  \propto B$'' when $A$ is proportional to $B$, that is, they are not
  necessarily asymptotic relations.}. Typical origins of such power-law
waiting times are complex energy landscapes \cite{xu11, saxton07, machta85,
  bouchaud90, miyaguchi11b} and diffusion in inner degrees of freedom
\cite{weiss86, goychuk02}. Similarly, EAMSD of GLE shows slow diffusion if
the memory kernel $\eta (t)$ decays algebraically $\eta(t) \sim
1/t^{\alpha}$ with $0<\alpha<1$ \cite{pottier03, vinales06}. A possible
origin of this non-Markovian memory effect is the viscoelasticity of the
medium \cite{mason95, *levine00}.


A primary difference between these two stochastic processes---CTRWs and
GLE---is in their ergodic properties. Even if the memory kernel of a GLE is
given by a power function as stated above, the GLE satisfies the ergodic
property, that is, time-averaged quantities coincide with ensemble-averaged
quantities \cite{deng09, *jeon10, *goychuk09, magdziarz11a}. In contrast,
this equivalence between the two averages does not hold in the CTRWs with
power-law waiting times. However, such CTRWs exhibit an extended form of
ergodicity---time-averaged quantities become random variables following a
distribution function \cite{miyaguchi11b, miyaguchi11c}.  This
distributional ergodicity is called weak ergodicity breaking
\cite{bouchaud92, korabel10, he08, lubelski08, miyaguchi11b, miyaguchi11c}
or infinite ergodicity \cite{aaronson97, akimoto10} .


In addition to the above mentioned examples, CTRWs are frequently used in
various fields of science \cite{scher75, young89, bouchaud90,
  song10}. However, there are cases in which some finite-size effects
should be considered in order to compare the model with experimental data.
In particular, cutoffs at the tail of the power-law waiting time
distribution \cite{xu11, saxton07, song10} and spatial confinement effects
\cite{jeon11, burov11} arise in many systems. The typical origins of
waiting time cutoffs are physical limits with respect to energy and spatial
extensions (e.g., \cite{xu11, saxton07, bouchaud90,
  matthews02}). Theoretical analysis of the CTRWs with such cutoffs is
difficult, because it is necessary to investigate the transient
behavior. However, this difficulty is avoidable by using the tempered
stable distribution (TSD) $P_{\rm TL} (\tau, \lambda)$ \cite{gajda10,
  *gajda11, magdziarz11a, miyaguchi11c}. The TSD has exponentially smooth
cutoffs, and is a modified version of the truncated stable distribution,
which has a sharp cutoff at the tail \cite{mantegna94}. Furthermore, the
TSD has the property called infinite divisibility \cite{koponen95, nakao00,
  cartea07, *del-Castillo-Negrete09, *stanislavsky09, *stanislavsky11,
  gajda10, *gajda11, magdziarz11a}, which allows a rigorous analysis even
for transient behavior. The TSD is a special model for the cutoffs, but it
shows typical behavior of cutoff distributions such as a slow convergence
to the Gaussian distribution \cite{koponen95, magdziarz11a, nakao00,
  cartea07, *del-Castillo-Negrete09, *stanislavsky09, *stanislavsky11,
  gajda10, *gajda11}. Furthermore, the generalized fractional Fokker-Planck
equation (GFFPE), which we will use in this paper to study EAMSD, has been
derived for CTRWs with tempered stable waiting times \cite{gajda10,
  *gajda11}. The ergodic properties of the system without confinement have
also been clarified in \cite{miyaguchi11c}, where the existence of a clear
transition from weak ergodicity breaking (an irreproducible regime) to
ordinary ergodicity (a reproducible regime) was shown analytically for the
case of a non-equilibrium initial ensemble (See Sec.~\ref{sec:TSD} for a
precise definition of the non-equilibrium and equilibrium ensembles).



Another important finite-size effect is spatial confinements. For example,
to understand the transport phenomena in cells \cite{golding06},
confinements due to cell membranes should be considered. Confinement
effects in CTRWs with power-law waiting times were studied numerically in
\cite{he08} and analytically in \cite{neusius09}; it was found that the
time-averaged MSD (TAMSD) shows a crossover from normal diffusion at short
timescales to anomalous slow diffusion at longer timescales. They also
reported numerically the weak ergodicity breaking in TAMSD.


Recently, the CTRW model with the two finite-size effects (i.e., waiting
time cutoffs and spatial confinements) has been used as a model for the
transport of lipid granules in living fission yeast cells \cite{jeon11,
  burov11}, in which confinement effects are caused by a Hookean force
exerted by optical tweezers. The model clearly explains the experimental
results such as the crossover in TAMSD and weak ergodicity breaking. These
studies mainly used numerical simulations, but a detailed theoretical
analysis has not yet been reported. Also, they studied only non-equilibrium
ensemble, and thus the dependences of the ergodic properties on initial
ensembles are still unknown.


In this paper, we present theoretical results for CTRWs with two
finite-size effects: the effects of cutoffs in the waiting time
distribution and the confinement effects. In particular, we focus on TAMSD
as an observable, and study the crossover from normal to anomalous
diffusion in the TAMSD and ergodic properties in terms of the scatter of
diffusion constant for the TAMSD. The TAMSD, which is often used in
single-molecule tracking experiments \cite{jeon11, burov11, golding06,
  bronstein09, graneli06, wang06, kusumi93, mason95, *levine00}, is defined
as
\begin{equation}
  \label{e.tamsd.1}
  \overline{(\delta x)^2} (\Delta,t)
  \equiv
  \frac {1}{t - \Delta}
  \int_0^{t-\Delta}
  | x (t'+\Delta)- x (t') |^2 dt',~~~
\end{equation}
where $x(t')$ is the position of the particle at time $t'$, $t$ is the
total measurement time, and $\Delta$ is the time interval. Hereinafter we
assume that $\Delta \ll t$. Here we define a generalized diffusion constant
$D$ as $\overline{(\delta x)^2} (\Delta,t) \simeq D \Delta^{\beta}$.  In
some experiments, it has been reported that $D$ behaves like a random
variable depending on each time series \cite{jeon11, burov11, golding06,
  bronstein09, graneli06}. Therefore, the scatter in TAMSD or $D$ has been
used to check the consistency of the model with experimental data
\cite{jeon11, burov11, jeon10a}.


In this paper, we use the TSDs \cite{koponen95, magdziarz11a, nakao00,
  cartea07, *del-Castillo-Negrete09, *stanislavsky09, *stanislavsky11,
  gajda10} as waiting time distributions of CTRWs. For the TSD, it is
possible to explicitly write the convoluted waiting distributions of any
order [Eq.~(\ref{e.prob.na})]. Moreover, we use the numerical method for
the TSD presented in \cite{gajda10} and Appendix \ref{sec:num.method}, and
study the initial and boundary value problem of the GFFPE to understand the
confinement effect.

The rest of this paper is organized as follows. In Sec.~\ref{sec:TSD}, we
introduce the TSDs. Then, in Sec.~\ref{sec:EA.TAMSD}, we show the crossover
from normal to anomalous diffusion in TAMSD by using the GFFPE. In
Sec.~\ref{sec:STAT.TAMSD}, ergodic properties of TAMSD are studied using
renewal theoretic analysis. Secs.~\ref{sec:conclusion} and
\ref{sec:discussion} are devoted to conclusion and discussion. In the
appendices, we summarize some technical matters, including a derivation of
GFFPE from CTRWs as well as the ergodic properties of general observables.

\section {Tempered Stable Distribution \label{sec:TSD}}

In this paper, we consider CTRWs confined in a one-dimensional lattice with
unit lattice constant: $(1,2,...,L)$. Jumps are permitted only to the
nearest neighbor sites without preferences, although this can be
generalized to jump length distributions with zero mean and finite
variances. It is also assumed that the particle is reflected if it goes
beyond the permitted region (for example, if the particle jumps into the
site $L+1$, it is pushed back to $L$). This system is a continuous-time
version of the discrete-time random walks (DTRWs) with reflecting barriers
\cite{feller68c16,*kemeny83}.

  
Furthermore, successive waiting times of the particle $\tau_k~(k=1,2,...)$
between jumps are assumed to be mutually independent and follow the TSD
$P_{\rm TL} (\tau_k, \lambda)$. The Laplace transform of the TSD,
$\tilde{P}_{\rm TL} (s, \lambda) \equiv \int_{0}^{\infty} d\tau e^{-s\tau}
P_{\rm TL} (\tau, \lambda)$, is given by
\begin{equation}
  \tilde{P}_{\rm TL} (s, \lambda) =
  \exp \left(-c \left[ (\lambda + s)^{\alpha} - \lambda^{\alpha} \right] \right),
  \label{e.laplace.of.ptl}
\end{equation}
where $\alpha \in (0, 1)$ is the stable index, $c$ is a scale factor, and
$\lambda \geq 0$ is a parameter characterizing the smooth cutoff. Note that
when $\lambda = 0$, this is the Laplace transform of the one-sided stable
distributions. Equivalently, the characteristic function of the TSD, $e^{\psi
  (s, \lambda)} \equiv \int_{-\infty}^{\infty} d\tau P_{\rm TL} (\tau,
\lambda) e^{i\zeta \tau} $, is given by
\begin{equation}
  \label{e.cfunc}
  e^{\psi (\zeta, \lambda)}
  =
  \exp \left(
  -c \left[ (\lambda - i\zeta)^{\alpha} - \lambda^{\alpha}
  \right] \right),
\end{equation}
This characteristic function is a special case of the TSD given in
\cite{koponen95, magdziarz11a, nakao00, cartea07, *del-Castillo-Negrete09,
  *stanislavsky09, *stanislavsky11, gajda10}. This is because $\tau$ takes
only positive values, and we need only one-sided distributions.
More precise definition of the TSD and the derivations of the above
equations are presented in Appendix \ref{sec:app.TSD}.  


The TSD in real space is also derived explicitly as follows (for a
derivation, see Appendix \ref{sec:app.TSD}):
\begin{eqnarray}
  \label{e.prob.1b}
  P_{\rm TL} (\tau, \lambda) &=&
  - \frac {e^{ c \lambda^{\alpha} - \lambda \tau}}{\pi \tau}
  \notag\\
  &&\,\times
  \sum_{k=1}^{\infty} \frac {\Gamma (k \alpha + 1)}{k!}
  \left(-c {\tau}^{-\alpha}\right)^k \sin (\pi k \alpha),
\end{eqnarray}
where $\Gamma (x)$ is the gamma function. When $\lambda=0$, $P_{\rm TL}
(\tau, 0)$ is the one-sided $\alpha$-stable distribution with a power-law
tail: $P_{\rm TL} (\tau, 0) \sim 1 / \tau^{1+\alpha}$ as $\tau \to \infty$
\cite{levy37, feller71c17}. 
Therefore, the TSD [Eq.~(\ref{e.prob.1b})] is the one-sided stable
distribution multiplied by the exponential function $e^{-\lambda \tau}$:
$P_{\rm TL} (\tau, \lambda) \propto e^{-\lambda \tau} P_{\rm TL} (\tau,
0)$. Thus, $P_{\rm TL} (\tau, \lambda)$ behaves as $P_{\rm TL} (\tau,
\lambda) \sim e^{-\lambda \tau} / \tau^{1+\alpha}$ when $\tau \to \infty$.

Moreover, the $n$-times convoluted PDF $P_{\rm TL}^{n} (\tau, \lambda)
$, which is the PDF of the sum of the successive waiting times $t_n \equiv
\sum_{k=1}^{n} \tau_k$, is expressed by using $P_{\rm TL}(\tau, \lambda)$:
\begin{equation}
  \label{e.prob.na}
  P_{\rm TL}^{n} (\tau, \lambda)
  =
  n^{- 1/ \alpha} P_{\rm TL} (n^{-1/\alpha} \tau, n^{1/\alpha} \lambda).
\end{equation}
Therefore, the $n$-times convoluted PDF $P_{\rm TL}^{n} (\tau, \lambda)$ is
also explicitly derived from Eqs.~(\ref{e.prob.1b}) and (\ref{e.prob.na}).
Using Eq.~(\ref{e.prob.na}), we derive transient properties of CTRWs,
including various crossovers, in the following sections.


Even though we use the TSD $P_{\rm TL} (\tau, \lambda)$ as a waiting time
distribution, we should further specify the first waiting time $\tau_1$, or
equivalently, the initial ensemble (Here, the first waiting time $\tau_1$
is the time interval between the start of the measurement and the first
jump. We always assume that measurements start at $t'=0$.). In this paper,
we consider two kinds of initial ensembles. The first one is a typical and
most frequently used non-equilibrium ensemble, for which the first waiting
times $\tau_1$ of the particles are chosen from $P_{\rm TL} (\tau,
\lambda)$. The second one is the equilibrium ensemble, for which the first
waiting times $\tau_1$ are chosen from the equilibrium waiting time
distribution $P_{\rm TL}^{\mathrm{eq}} (\tau, \lambda)$. Here, $P_{\rm
  TL}^{\mathrm{eq}} (\tau, \lambda)$ can be defined by its Laplace
transformation \cite{cox62, godrche01}:
\begin{equation}
  \tilde{P}_{\rm TL}^{\mathrm{eq}} (s, \lambda) =
  \frac {1 - \tilde{P}_{\rm TL} (s, \lambda)}
  { \left\langle \tau \right\rangle s},
  \label{e.laplace.of.ptl.eq}
\end{equation}
where $\left\langle \tau \right\rangle = c\lambda^{\alpha-1} \alpha$ is the
mean waiting time for $P_{\rm TL} (\tau, \lambda)$. Note that the second
and subsequent waiting times, $\tau_2, \tau_3, \dots$, are chosen from
$P_{\rm TL} (\tau, \lambda)$ for both ensembles. Numerical methods to
generate random variables following $P_{\rm TL} (\tau, \lambda)$ and
$P_{\rm TL}^{\mathrm{eq}} (\tau, \lambda)$ are summarized in Appendix
\ref{sec:num.method}. $P_{\rm TL}^{\mathrm{eq}} (\tau, \lambda)$ can be
expressed analytically as follows

\begin{align}
  \label{e.prob.1b.eq}
  P_{\rm TL}^{\mathrm{eq}} (\tau, \lambda) &=
  - \frac {e^{ c \lambda^{\alpha}}}{\pi \left\langle \tau \right\rangle}
  \notag\\
  & \times
  \sum_{k=1}^{\infty} \frac {\Gamma (k \alpha + 1)}{k!}
  \left(-c {\tau}^{-\alpha}\right)^k \sin (\pi k \alpha) f_k(\tau),
\end{align}
where $f_k(\tau)$ is defined by $f_k(\tau) = \int_{0}^{1} e^{-\lambda \tau
  /a} a^{\alpha k - 1} da$. See Appendix \ref{sec:app.TSD} for a derivation
of Eq.~(\ref{e.prob.1b.eq}). The mean waiting time for $P_{\rm
  TL}^{\mathrm{eq}} (\tau, \lambda)$ is given by $\left\langle \tau
\right\rangle_{\mathrm{eq}} = (c\lambda^{\alpha}\alpha + 1 - \alpha)/(2
\lambda) \simeq (1 - \alpha)/(2 \lambda) $, which is much longer than
$\left\langle \tau \right\rangle$ if $\lambda$ is small.

\section {Ensemble Average of TAMSD \label{sec:EA.TAMSD}}

The confinement effects on CTRWs with power-law waiting times were
investigated in \cite{neusius09} by using the Fractional Fokker-Planck
equation (FFPE). Here instead of the FFPE, we use GFFPE, which was derived
in \cite{gajda10}, to incorporate the smooth cutoff into the waiting time
distribution and to study the ensemble averages of the TAMSD.


There are three timescales in the present model: (1) time interval
$\Delta$, (2) total measurement time $t$, and (3) timescale of the cutoff
$1/\lambda$. Let us define the Laplace variables $u$ and $s$ conjugate to
$\Delta$ and $t$, respectively. Since we assume that $\Delta \ll t$ ($s \ll
u$), it is sufficient to consider the following three cases:
\begin{eqnarray}
  \label{e.confine.timescales.2}
  \begin{array}{clcl}
    \mathrm{(A)} & \Delta \ll t \ll 1/\lambda &~~\Longleftrightarrow~~& \lambda \ll s \ll u,
    \\[.2cm]
    \mathrm{(B)} & \Delta \ll 1/\lambda \ll t &~~\Longleftrightarrow~~& s \ll \lambda \ll u,
    \\[.2cm]
    \mathrm{(C)} & 1/\lambda \ll \Delta \ll t &~~\Longleftrightarrow~~& s \ll u \ll \lambda.
  \end{array}
\end{eqnarray}
Here it is expected that the standard random walk behavior arises in case
(C). Thus, we only study the cases (A) and (B) in this paper.

\subsection {Decomposition of ensemble average of TAMSD}

In this subsection, we rewrite the ensemble averages of TAMSD $\langle
\overline{(\delta x)^2} (\Delta,t) \rangle$ and $\langle \overline{(\delta
  x)^2} (\Delta,t) \rangle_{\mathrm{eq}}$ by using the EAMSD $\left\langle
|x (\Delta) - x(0)|^2\right\rangle$ [see
Eqs.~(\ref{e.confine.ea-tamsd.laplace}) and
(\ref{e.confine.ea-tamsd.laplace.eq}) ]. Here and in the followings, we use
the bracket $\left\langle \cdot \right\rangle$ for the average over the
non-equilibrium initial ensemble, while $\left\langle \cdot
\right\rangle_{\mathrm{eq}}$ for the average over the equilibrium initial
ensemble. Also, we assume that initial position $x(0)$ is uniformly
distributed on the lattice $\{1, \dots, L\}$ for both ensembles.

First, let $w(\tau)$ be the waiting time distribution of a renewal process
with a finite mean, $\left\langle \tau \right\rangle < \infty$. Moreover,
we define $w_e (\tau; t')$ as the PDF of the forward recurrence time $\tau$
\cite{godrche01}, i.e., $w_e (\tau; t')$ is the waiting time distribution
at time $t'$. Here, note that $t'$ is not necessarily a renewal
time. Particularly, $w_{\mathrm{e}}(\tau; 0) = w(\tau)$ for the
non-equilibrium initial ensemble, since we assume that we start
measurements at $t'=0$, whereas $w_{\mathrm{e}}(\tau; t') \neq
w(\tau)$ in general. By contrast, for the equilibrium ensemble,
\begin{equation}
  \label{e.time_translation_inv}
  w_{\mathrm{e}}^{\mathrm{eq}} (\tau; t')
  =
  w^{\mathrm{eq}}(\tau),
\end{equation}
where $w_{\mathrm{e}}^{\mathrm{eq}} (\tau; t')$ is the PDF of the forward
recurrence time at time $t'$ for the equilibrium ensemble, and
$w^{\mathrm{eq}}(\tau)$ is defined through its Laplace transformation
$\tilde{w}^{\mathrm{eq}}(s) \equiv \{1 - \tilde{w}
(s)\} / \left\langle \tau \right\rangle s$ [See
Eq.~(\ref{e.laplace.of.ptl.eq})].  This relation
[Eq.~(\ref{e.time_translation_inv})] is obvious because of the
time-translation invariance of the equilibrium state, and a proof is
given in Appendix \ref{sec:app.time.trans}.

\subsubsection {Non-equilibrium ensemble}

Let us begin with the non-equilibrium ensemble. Using the PDF
$w_{\mathrm{e}}(\tau; t')$, we can express the ensemble average of the
displacement during $[t', t'+\Delta]$ as follows:
\begin{eqnarray}
  &&\left\langle [ x(t'+\Delta)-x(t')]^2 \right\rangle
  \notag\\[0.1cm]
  &&\quad=
  \sum_{l = \pm 1}^{}\frac {1}{2}
  \int_0^{\Delta} d\tau w_{\mathrm{e}}(\tau; t')
  \left\langle
  \left|x(t'+\Delta) - x(t'+\tau) + l\right|^2
  \right\rangle
  \notag\\[0.1cm]
  \label{e.confine.frec_time}
  &&\quad=
  \int_0^{\Delta} d\tau w_{\mathrm{e}}(\tau; t')
  \left[\left\langle
  \left|x(\Delta - \tau) - x(0)\right|^2
  \right\rangle + 1\right].~~
\end{eqnarray}
This is an exact relation, while an approximated version of this equation
is already presented in Ref.\cite{neusius09}. We also show a more detailed
derivation in Appendix \ref{sec:app.eq10}.
From Eqs.~(\ref{e.tamsd.1}) and (\ref{e.confine.frec_time}), we have
\begin{eqnarray}
  &&\langle \overline{(\delta x)^2} (\Delta,t) \rangle\notag\\[0.1cm]
  &&\quad=
  \int_0^{\Delta} d\tau \bar{w}_{\mathrm{e}}(\tau; t)
  \left[
  \left\langle \left|x(\Delta - \tau) - x(0)\right|^2 \right\rangle + 1
  \right],
  \qquad
  \label{e.confine.ea-tamsd}
\end{eqnarray}
where $\bar{w}_{\mathrm{e}}(\tau; t)$ is defined as
\begin{equation}
  \label{e.confine.w_bar}
  \bar{w}_{\mathrm{e}}(\tau; t)
  \equiv
  \int_0^{t-\Delta} dt' \frac {w_{\mathrm{e}}(\tau; t')}{t - \Delta}.
\end{equation}
From Eq.~(\ref{e.confine.ea-tamsd}), the ensemble average of TAMSD is given
by the convolution of $\bar{w}_{\mathrm{e}}(\tau; t)$ and the EAMSD
$\langle |x(\Delta - \tau) - x(0)|^2 \rangle$.  In this and the following
subsections, we study these two factors using Laplace transformations. Here
we further rewrite Eqs.~(\ref{e.confine.ea-tamsd}) and
(\ref{e.confine.w_bar}) using Laplace transformations. First, the Laplace
transform of Eq.~(\ref{e.confine.w_bar}) with respect to $\tau$ gives
\begin{equation}
  \label{e.confine.w_bar.laplace}
  \tilde{\bar{w}}_{\mathrm{e}}(u; t)
  =
  \int_0^{t-\Delta} dt' \frac {\tilde{w}_{\mathrm{e}}(u; t')}{t - \Delta},
\end{equation}
where we have defined the Laplace transformations of
${w}_{\mathrm{e}}(\tau; t)$ and $\bar{w}_{\mathrm{e}}(\tau; t)$ as %
$\tilde{w}_{\mathrm{e}}(u; t) \equiv \int_{0}^{\infty} w_{\mathrm{e}}(\tau;
t) e^{-u \tau}d\tau$ %
and %
$\tilde{\bar{w}}_{\mathrm{e}}(u; t) \equiv \int_{0}^{\infty}
\bar{w}_{\mathrm{e}}(\tau; t) e^{-u \tau}d\tau$, respectively. Furthermore,
the Laplace transform of Eq.~(\ref{e.confine.w_bar.laplace}) with respect
to $t$ gives
\begin{equation}
  \breve{\bar{w}}_{\mathrm{e}}(u; s)
  \label{e.confine.w_bar.laplace.2}
  =
  e^{- s\Delta}
  \int_s^{\infty} ds'
  \frac {\breve{w}_{\mathrm{e}}(u; s')}{s'},
\end{equation}
where the double Laplace transformations $\breve{w}_{\mathrm{e}}(u; s)$ and
$\breve{\bar{w}}_{\mathrm{e}}(u; s)$ are defined as
$\breve{w}_{\mathrm{e}}(u; s) \equiv \int_{0}^{\infty}
\tilde{w}_{\mathrm{e}}(u; t) e^{-st}dt$ and
$\breve{\bar{w}}_{\mathrm{e}}(u; s) \equiv \int_{\Delta}^{\infty}
\tilde{\bar{w}}_{\mathrm{e}}(u; t) e^{-st}dt$, respectively. In addition,
by taking the Laplace transformation of Eq.~(\ref{e.confine.ea-tamsd}) with
respect to $\Delta$, we obtain
\begin{eqnarray}
  &&\mathcal{L}
  \left[
  \langle \overline{(\delta x)^2} (\Delta,t) \rangle
  \right] (u, t)
  \notag\\[0.1cm]
  \label{e.confine.ea-tamsd.laplace}
  &&\quad=
  \tilde{\bar{w}}_{\mathrm{e}}(u; t)
  \left\{
  \,\mathcal{L}
  \left[ \left\langle \left|x(\Delta) - x(0)\right|^2 \right\rangle \right] (u)
  + \frac {1}{u}
  \right\}.
\end{eqnarray}

For the non-equilibrium ensemble, the following relation between the
waiting time distribution $w(t)$ and forward recurrence time distribution
$w_{\mathrm{e}} (\tau; t')$ is well known \cite{cox62,godrche01, barkai03a,
  *margolin04}:
\begin{equation}
  \label{e.confine.w_e}
  \breve{w}_{\mathrm{e}} (u;s)
  =
  \frac {\tilde{w}(u) - \tilde{w}(s)}{s-u} \frac {1}{1 - \tilde{w}(s)}.
\end{equation}
See Appendix \ref{sec:app.time.trans} for a derivation.  If we choose
TSD [Eq.~(\ref{e.prob.1b})] for the waiting time distribution $w(\tau)$, we
have
\begin{equation}
  \label{e.confine.w_e.tlf}
  \breve{w}_{\mathrm{e}} (u;s)
  \simeq
  \frac {(\lambda + u)^{\alpha} - (\lambda + s)^{\alpha}}{u-s}
  \frac {1}{(\lambda + s)^{\alpha} - \lambda^{\alpha}},
\end{equation}
where we have used Eq.~(\ref{e.laplace.of.ptl}) and $\lambda, u, s \ll
1$. Using the Eq.~(\ref{e.confine.w_e.tlf}) and $s \ll u$, the integral on
the RHS of Eq.~(\ref{e.confine.w_bar.laplace.2}) can be approximated as
\begin{eqnarray}
  \label{e.confine.w_e.integral}
  \int_s^{\infty} ds' \frac {\breve{w}_{\mathrm{e}}(u; s')}{s'}
  &\simeq&
  \left\{
  \begin{array}{ll}
    \displaystyle \frac {u^{\alpha-1}}{\alpha s^{\alpha} },
    &~\mathrm{for}~~\lambda \ll s
    \\[0.5cm]
    \displaystyle
    \frac {\lambda^{1-\alpha} u^{\alpha-1}}{\alpha s },
    &~\mathrm{for}~~s \ll \lambda
  \end{array}
  \right.
\end{eqnarray}
Then, the inverse Laplace transformation of
Eq.~(\ref{e.confine.w_bar.laplace.2}) with respect to $s$ gives
\begin{eqnarray}
  \label{e.confine.bar_w_e}
  \tilde{\bar{w}}_{\mathrm{e}}(u; t)
  &\simeq&
  \left\{
  \begin{array}{ll}
    \displaystyle \frac {u^{\alpha-1}}{\Gamma (\alpha +1)} t^{\alpha -1} ,
    &~\mathrm{for}~~t \ll 1/\lambda
    \\[0.5cm]
    \displaystyle
    \frac {\lambda^{1-\alpha} u^{\alpha-1}}{\alpha},
    &~\mathrm{for}~~1/\lambda \ll t
  \end{array}
  \right.
\end{eqnarray}
where we have used $e^{-s\Delta} \simeq 1$ [Note that $s \Delta \ll 1$
because of Eq.~(\ref{e.confine.timescales.2})].

\subsubsection {Equilibrium ensemble}

For the case of the equilibrium ensemble, Eq.~(\ref{e.confine.frec_time})
should be replaced by
  \begin{eqnarray}
  &&\left\langle \left[ x(t'+\Delta)-x(t')\right]^2 \right\rangle_{\mathrm{eq}}
  \notag\\[0.1cm]
  \label{e.confine.frec_time.eq}
  &&\qquad=
  \int_0^{\Delta} d\tau w^{\mathrm{eq}}(\tau)
  \left[
  \left\langle \left|x(\Delta - \tau) - x(0) \right|^2\right\rangle
  + 1
  \right].~~
\end{eqnarray}
Note that the ensemble average in the right-hand side (RHS) is taken over
the non-equilibrium ensemble. Since RHS of the above equation is
independent of $t'$, we have the following equation through the calculation
similar to that in the non-equilibrium case:
  \begin{eqnarray}
  &&\mathcal{L}
  \left[
  \langle \overline{(\delta x)^2} (\Delta) \rangle_{\mathrm{eq}}
  \right] (u) \notag\\[0.1cm]
  \label{e.confine.ea-tamsd.laplace.eq}
  &&\qquad=
  \tilde{w}^{\mathrm{eq}}(u)
  \left\{
  \mathcal{L}
  \left[ \left\langle \left|x(\Delta) - x(0) \right|^2\right\rangle \right] (u)
  + \frac {1}{u}
  \right\}.\quad
\end{eqnarray}
Moreover, the left-hand-side (LHS) of Eq.~(\ref{e.confine.frec_time.eq})
with $t'=0$ is just the EAMSD with respect to the equilibrium initial
ensemble. Thus, using Eq.~(\ref{e.tamsd.1}) we have 
\begin{equation}
  \left\langle [ x(\Delta)-x(0)]^2 \right\rangle_{\mathrm{eq}}
  =
  \langle \overline{(\delta x)^2} (\Delta) \rangle_{\mathrm{eq}}
  \label{e.confine.ergode.eq}
\end{equation}
This is a manifestation of ergodicity for the equilibrium ensemble. Note
that Eq.~(\ref{e.confine.ergode.eq}) is valid in general, if $\left\langle
\tau \right\rangle$ is finite (and thus $w^{\mathrm{eq}}(\tau)$ exists).

If we choose TSD [Eqs.~(\ref{e.prob.1b}) and (\ref{e.laplace.of.ptl})] for
the waiting time distribution $w(\tau)$, we have the following relation for
$\tilde{w}^{\mathrm{eq}}(u)$ \cite{godrche01}:
\begin{equation}
  \label{e.confine.w_e.tlf.eq}
  \tilde{w}^{\mathrm{eq}}(u) =
  \frac {1 - e^{-c\left[ (\lambda + u)^{\alpha} - \lambda^{\alpha} \right]}}
  {\left\langle \tau \right\rangle u}
  \simeq
  \frac {1}{\alpha} \left( \frac {\lambda}{u}\right)^{1 - \alpha},
\end{equation}
where we used the assumption $\lambda \ll u$
[Eq.~(\ref{e.confine.timescales.2})] and $\left\langle \tau \right\rangle =
c\lambda^{\alpha-1} \alpha$.

\subsection {Ensemble average of MSD}
Furthermore, we have to calculate the Laplace transform of EAMSD
$\left\langle |x (\Delta) - x(0)|^2\right\rangle$ in order to obtain the
ensemble average of TAMSD [Eqs.~(\ref{e.confine.ea-tamsd.laplace}) and
(\ref{e.confine.ea-tamsd.laplace.eq})]. First, let us express the EAMSD
approximately as follows \cite{neusius09}:
\begin{equation}
  \label{e.confine.eamsd}
  \left\langle |x(\Delta) - x(0)|^{2} \right\rangle
  \simeq
  \int_{0}^{L} \frac {dx_s}{L} \int_{0}^{L} dx
  P(x,\Delta; x_s, 0) (x-x_s)^{2},
\end{equation}
where $P(x,\tau; x_s, 0)$ is the transition probability from $x_s ~(t=0)$
to $x ~(t=\Delta)$ of the GFFPE (See Appendix \ref{sec:appd}).  The above
equation is correct in the hydrodynamic limit $L \to \infty$.  Also, it
should be noted that the GFFPE is an equation for the non-equilibrium
ensemble, and therefore the EAMSD given by Eq.~(\ref{e.confine.eamsd}) is
also for the non-equilibrium ensemble. The Laplace transformation of
Eq.~(\ref{e.confine.eamsd}) is given by
\begin{eqnarray}
  \mathcal{L}
  &&\left[  \left\langle |x(\Delta) - x(0)|^{2} \right\rangle \right] (u)
  \notag\\[0.1cm]
  &&=
  \int_{0}^{L} \frac {dx_s}{L} \int_{0}^{L} dx
  \tilde{P}(x,u; x_s, 0) (x-x_s)^{2}
  \notag\\[0.2cm]
  &&=
  \frac {L^2}{6}
  \Bigg[
  \frac {1}{u} -
  \frac {96}{\pi^{4}}
  {\sum_{\begin{subarray}{c}n=1 \\ n: \mathrm{odd}\end{subarray} }^{\infty} }
  \frac {n^{-4}}
  {u + (n\pi / L)^2 K u\tilde{M}(u)}
  \Bigg]
  \notag \\[0.2cm]
  \label{e.confine.eamsd.laplace.2}
  &&\simeq
  \frac {L^2}{6u}
  \Bigg[
  1  -
  \frac {96}{\pi^{4}}
  \sum_{\begin{subarray}{c}n=1 \\ n: \mathrm{odd}\end{subarray} }^{\infty}
  \frac 1{n^{4}}
  \frac {(\Delta_c u)^{\alpha}}{(\Delta_c u)^{\alpha} + n^{2}}
  \Bigg],~~~~~
\end{eqnarray}
where we use $\lambda \ll u ~(\Delta \ll 1/\lambda)$. Moreover, $K$ is
defined as $K \equiv 1/2c$, and '$n: \mathrm{odd}$' under the $\sum$ means
that the summation is taken over odd terms. We also define a characteristic
timescale $\Delta_c$ as
\begin{equation}
  \label{e.confine.eamsd.laplace.tauc}
  \Delta_c \equiv \left(\frac {2cL^{2}}{\pi^{2}}\right)^{1/\alpha}.
\end{equation}
Note that Eq.~(\ref{e.confine.eamsd.laplace.2}) is the same as the one
obtained in \cite{neusius09} for the case of the power-law waiting time.
This means that this property of EAMSD is independent of the timescale of
the smooth cutoff $1/\lambda$.


Finally, we obtain the following estimate for the Laplace transformation of
EAMSD:

\begin{align}
  \label{e.confine.eamsd.laplace.4}
  \mathcal{L} \left[  \left\langle |x (\Delta) - x(0)|^2\right\rangle \right] (u)
  \simeq
  \begin{cases}
    \dfrac {1}{cu^{1+\alpha}},
    &\mathrm{for}\, \Delta_c u \gg 1
    \\[0.35cm]
    \dfrac {L^2}{6u},
    &\mathrm{for}\, \Delta_c u \ll 1,
  \end{cases}
\end{align}
where we used the RHS of Eq.~(\ref{e.confine.eamsd.laplace.2}) for the case
of $\Delta_c u \ll 1$, while we rewrote the RHS of
Eq.~(\ref{e.confine.eamsd.laplace.2}) by using zeta functions (see Appendix
\ref{sec:appe-riemann}) for the case of $\Delta_c u \gg 1$ as
\begin{equation}
  \label{e.confine.eamsd.laplace.aux}
  \frac {16L^2}{\pi^{4} \Delta_c^{\alpha}  u^{1+\alpha} }
  \Bigg[
  \frac {\pi^{2}}{8}
  -
  \sum_{\begin{subarray}{c}n=1 \\ n: \mathrm{odd}\end{subarray} }^{\infty}
  \frac {1} {(\Delta_c  u )^{\alpha} + n^{2}}
  \Bigg].
\end{equation}
Then, the summation term can be neglected since $\Delta_c u \gg 1$.  From
Eq.~(\ref{e.confine.eamsd.laplace.4}), we obtain the EAMSD for
non-equilibrium ensemble as follows:

\begin{align}
  \label{e.confine.eamsd.5}
  \left\langle |x(\Delta) - x(0)|^2 \right\rangle
  \simeq
  \begin{cases}
    \dfrac {\Delta^{\alpha}}{c\Gamma (1+\alpha)},
    &\mathrm{for}\, \Delta \ll \Delta_c
    \\[0.35cm]
    \dfrac {L^2}{6},
    &\mathrm{for}\, \Delta \gg \Delta_c.
  \end{cases}
\end{align}
As shown in the next subsection, the TAMSD for the non-equilibrium ensemble
behaves differently. Namely, the ergodicity is broken even at $t \to
\infty$. In contrast, the ergodicity is satisfied for the equilibrium
ensemble at $t \to \infty$.

\subsection {Ensemble average of TAMSD}

In this subsection, we derive asymptotic behavior of the ensemble-averaged
TAMSD using the results from the preceding subsections.

\begin{figure}[t]
  \centerline{\includegraphics[width=8.2cm]{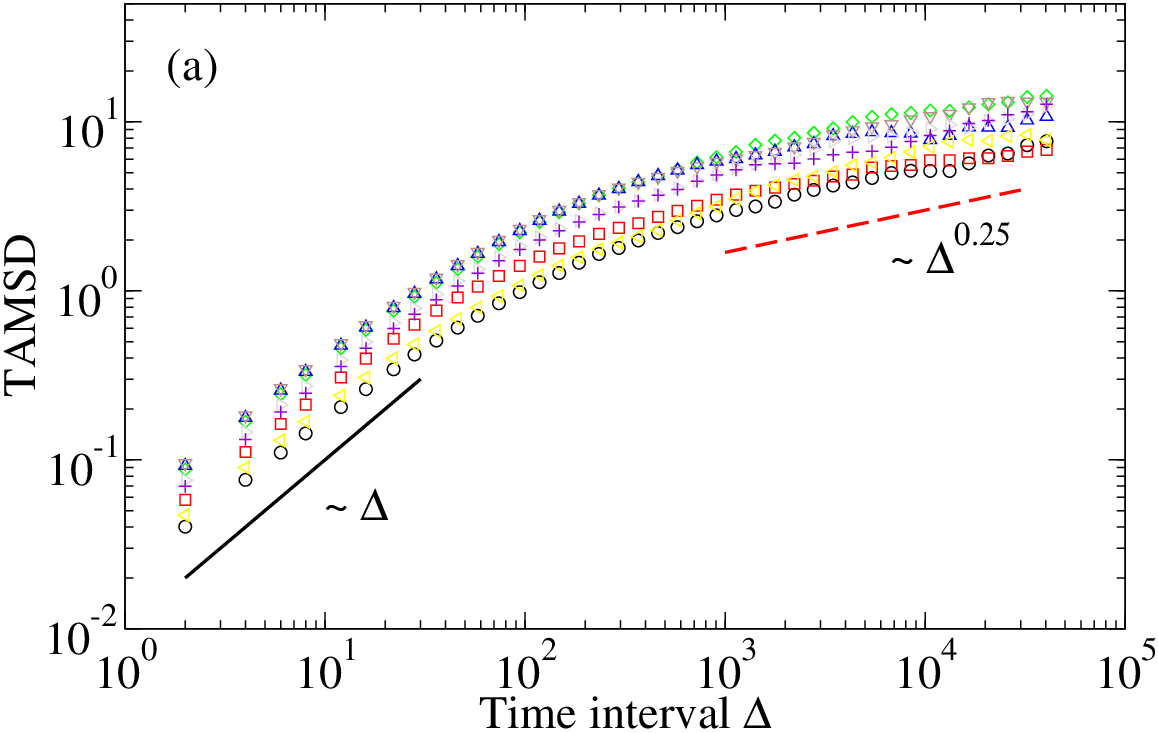}}
  \vspace*{.7cm}
  \centerline{\includegraphics[width=8.2cm]{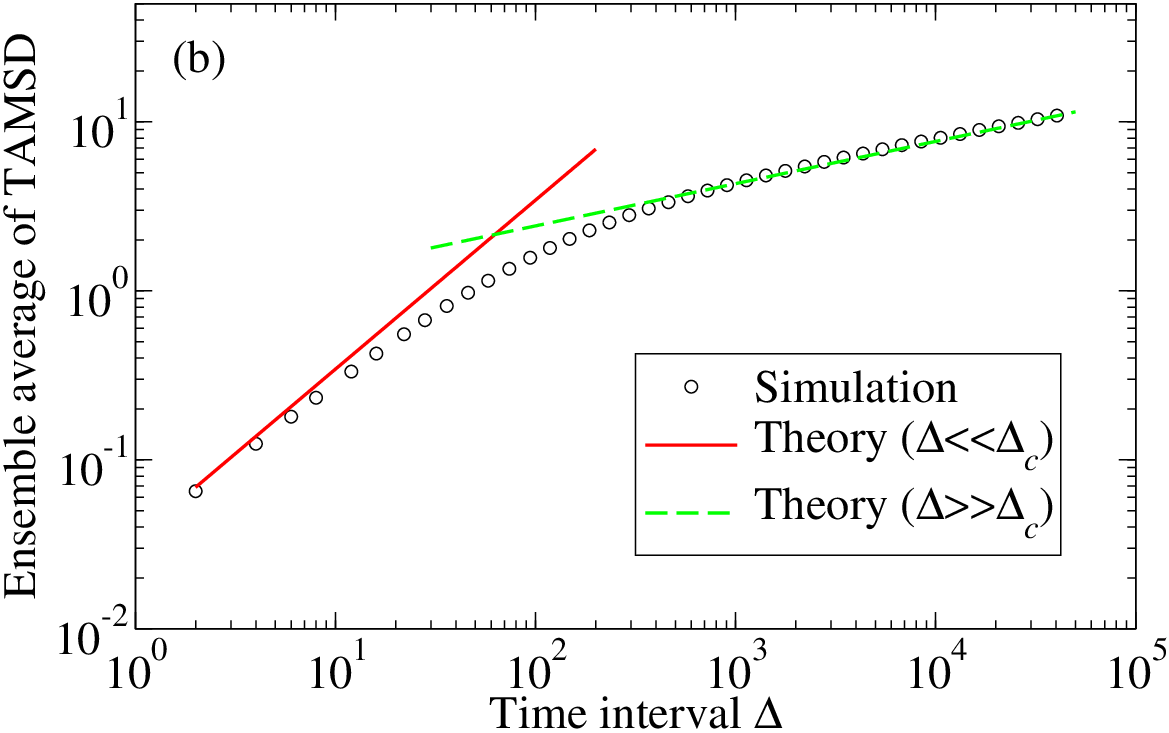}}
  \caption{\label{f.confine.msd.1} (Color online) (a) TAMSD
    $\overline{(\delta x)^2} (\Delta,t)$ vs. time interval $\Delta$ in
    log--log form for the non-equilibrium ensemble (A): $t<1/\lambda$. The
    total measurement time $t$ is set as $t=10^{6}$, and the cutoff
    parameter $\lambda$ as $\lambda=10^{-7}$. Other parameters are set as
    $\alpha=0.75, c=1$, and $L=11$. TAMSD is calculated for 8 different
    realizations of trajectories and different symbols correspond to
    different realizations. (b) The ensemble average of TAMSD in log--log
    form (circles). The lines are the theoretical predictions given by
    Eq.~(\ref{e.confine.ea-tamsd.final.1.real}). Note that no adjustable
    parameters were used to obtain these theoretical lines.  }
\end{figure}

\subsubsection {Non-equilibrium ensemble [case (A)]:\, $t \ll 1/\lambda$}

First, we start with the non-equilibrium ensemble for $t \ll
1/\lambda$. From Eqs.~(\ref{e.confine.ea-tamsd.laplace}),
(\ref{e.confine.bar_w_e}), and (\ref{e.confine.eamsd.laplace.4}), we have
leading terms in $u \ll 1$ and $L \gg 1$ as follows:

\begin{align}
  &\mathcal{L}
  \left[
  \langle \overline{(\delta x)^2} (\Delta,t) \rangle
  \right] (u,t)
  \notag\\[0.1cm]
  \label{e.confine.ea-tamsd.final.1}
  &\qquad=
  \begin{cases}
    \dfrac {u^{- 2}}{c\Gamma (1+\alpha) t^{1-\alpha}},
    &\mathrm{for}~ \Delta_c u \gg 1
    \\[0.4cm]
    \dfrac {L^{2}}{6}
    \dfrac {u^{\alpha-2}}{\Gamma (1+\alpha) t^{1 -\alpha}},
    &\mathrm{for}~ \Delta_c u \ll 1.
  \end{cases}
\end{align}
The inverse Laplace transformation with respect to $u$ gives
\begin{eqnarray}
  \label{e.confine.ea-tamsd.final.1.real}
  \langle \overline{(\delta x)^2} (\Delta,t) \rangle
  =
  \left\{
  \begin{array}{ll}
    \displaystyle
    \frac {\Delta}{c\Gamma (1+\alpha) t^{1-\alpha}},
    &\mathrm{for}~ \Delta \ll \Delta_c
    \\[0.4cm]
    \displaystyle
    \frac
    {L^{2}\Delta^{1 -\alpha}}
    {6\Gamma(1+\alpha) \Gamma (2 - \alpha) t^{1 -\alpha}},
    &\mathrm{for}~ \Delta_c \ll \Delta.
  \end{array}
  \right.
\end{eqnarray}
Thus, the ensemble-averaged TAMSD shows normal diffusion at a short
timescale ($\Delta \ll \Delta_c$) and anomalous slow diffusion at a longer
timescale ($\Delta_c \ll \Delta.$). As expected, these results perfectly
coincide with those of the previous studies [Eqs.~(10) and (11) in
\cite{neusius09}]. Also, in this regime, the TAMSD depends on the
measurement time $t$. That is, the diffusion becomes slower, as the
measurement time increases. We call this behavior \textit{aging} in this
article.

In Fig.~\ref{f.confine.msd.1}(a), TAMSDs for 8 different trajectories are
shown. Although these TAMSDs show similar scaling behavior, the diffusion
constant $D$ of each TAMSD $\overline{(\delta x)^2} (\Delta,t) \sim D
\Delta^{\gamma}$ seems randomly distributed. This behavior is analyzed in
the next section. In Fig.~\ref{f.confine.msd.1} (b), the ensemble-averaged
TAMSD $\langle \overline{(\delta x)^2} (\Delta,t) \rangle$ is
displayed. The solid and dashed lines are the theoretical predictions given
by Eq.~(\ref{e.confine.ea-tamsd.final.1.real}).

\begin{figure}[t]
  \centerline{\includegraphics[width=8.2cm]{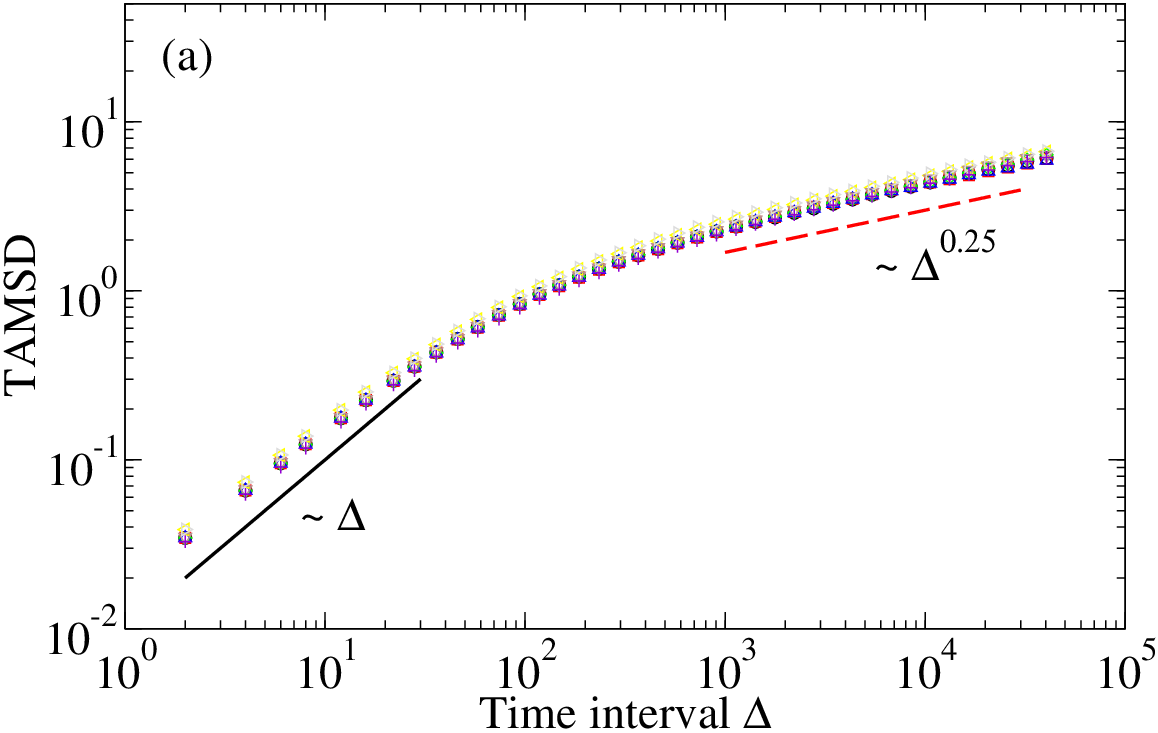}}
  \vspace*{.7cm}
  \centerline{\includegraphics[width=8.2cm]{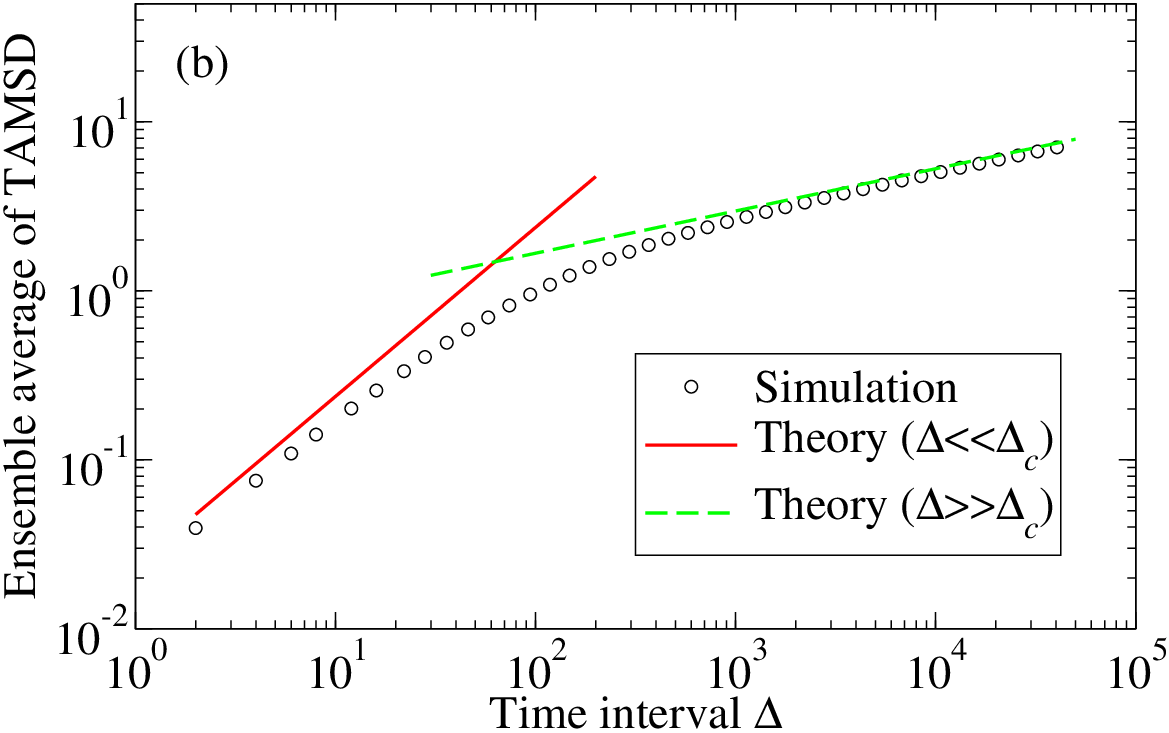}}
  \caption{\label{f.confine.msd.2} (Color online) (a) TAMSD
    $\overline{(\delta x)^2} (\Delta,t)$ vs. time interval $\Delta$ in
    log--log form for the non-equilibrium ensemble (B): $t >
    1/\lambda$. The total measurement time $t$ is set as $t=10^{8}$, and
    the cutoff parameter $\lambda$ as $\lambda=10^{-7}$. The other
    parameters are the same as in Fig.~\ref{f.confine.msd.1}. TAMSD is
    calculated for 8 different realizations of trajectories and different
    symbols correspond to different realizations. (b) The ensemble average
    of TAMSD in log--log form (circles). The lines are theoretical
    predictions given by Eq.~(\ref{e.confine.ea-tamsd.final.2.real}). No
    adjustable parameters were used to obtain these theoretical lines.%
  }
\end{figure}

\subsubsection {Non-equilibrium ensemble [case (B)]:\, $1/\lambda \ll t$}

Next, we study the case of $1/ \lambda \ll t$.  From
Eqs.~(\ref{e.confine.ea-tamsd.laplace}), (\ref{e.confine.bar_w_e}), and
(\ref{e.confine.eamsd.laplace.4}), we have
\begin{eqnarray}
  \label{e.confine.ea-tamsd.final.2}
  \mathcal{L}
  \left[
  \langle \overline{(\delta x)^2} (\Delta,t) \rangle
  \right] (u,t)
  =
  \left\{
  \begin{array}{ll}
    \displaystyle
    \frac {1}{c\lambda^{\alpha-1}\alpha}
    \frac {1}{u^{2}},
    &\mathrm{for}~ \Delta_c u \gg 1
    \\[0.5cm]
    \displaystyle
    \frac {1}{\lambda^{\alpha-1}\alpha} \,\frac {L^2}{6u^{2-\alpha}},
    &\mathrm{for}~ \Delta_c u \ll 1.
  \end{array}
  \right.
\end{eqnarray}
Then, taking the inverse Laplace transformation, we obtain
\begin{eqnarray}
  \label{e.confine.ea-tamsd.final.2.real}
  \langle \overline{(\delta x)^2} (\Delta,t) \rangle
  =
  \left\{
  \begin{array}{ll}
    \displaystyle
    \frac {\Delta}{c\lambda^{\alpha-1}\alpha},
    &\mathrm{for}~ \Delta \ll \Delta_c
    \\[0.5cm]
    \displaystyle
    \frac {1 }{\lambda^{\alpha-1} \alpha} \frac {L^2\Delta^{1 -\alpha}}{6\Gamma (2 - \alpha)},
    &\mathrm{for}~ \Delta_c \ll \Delta.
  \end{array}
  \right.
\end{eqnarray}
Thus, a crossover from the normal to anomalous diffusion similar to that in
case (A) can be observed even in the long measurement time limit $t \to
\infty$, whereas the aging behavior---$t$-dependence of the diffusion
constant---vanishes. Note also that even in this limit, $t \to \infty$, the
(ensemble-averaged) TAMSD does not coincide with the EAMSD
[Eq.~(\ref{e.confine.eamsd.5})].

Fig.~\ref{f.confine.msd.2}(a) shows TAMSDs for 8 different trajectories. In
this case, the scatter of TAMSDs, observed in
Fig.~\ref{f.confine.msd.1}(a), is diminished. The ensemble-averaged TAMSD
is also shown in Fig.~\ref{f.confine.msd.2}(b) by circles, where the
theoretical predictions given by
Eq.~(\ref{e.confine.ea-tamsd.final.2.real}) are shown by solid and dashed
lines.

\subsubsection {Equilibrium ensemble}

For the case of the equilibrium initial ensemble, we have the following
relation from Eqs.~(\ref{e.confine.ea-tamsd.laplace.eq}),
(\ref{e.confine.w_e.tlf.eq}), and (\ref{e.confine.eamsd.laplace.4}):

\begin{align}
  \label{e.confine.ea-tamsd.laplace.final.eq}
  \mathcal{L}
  \left[
  \langle \overline{(\delta x)^2} (\Delta) \rangle_{\mathrm{eq}}
  \right] (u)
  \simeq
  \begin{cases}
    \dfrac {1}{c \lambda^{\alpha-1} \alpha} \dfrac {1}{u^2},
    &\text{for}~ \Delta_c u \gg 1
    \\[0.5cm]
    \dfrac {1}{\lambda^{\alpha-1}\alpha} \,\dfrac {L^2}{6u^{2-\alpha}},
    &\text{for}~ \Delta_c u \ll 1.
  \end{cases}
\end{align}
Through the inverse transformation, we have

\begin{align}
  \label{e.confine.ea-tamsd.eq}
  \langle \overline{(\delta x)^2} (\Delta) \rangle_{\mathrm{eq}}
  \simeq
  \begin{cases}
    \dfrac {\Delta}{c\lambda^{\alpha-1}\alpha},
    &\text{for}~ \Delta \ll \Delta_c
    \\[0.5cm]
    \dfrac {1}{\lambda^{\alpha-1}\alpha}
    \dfrac {L^2 \Delta^{1-\alpha}}{6\Gamma(2-\alpha)},
    &\text{for}~ \Delta \gg \Delta_c.
  \end{cases}
\end{align}

The equation (\ref{e.confine.ea-tamsd.eq}) is equivalent to
Eq.~(\ref{e.confine.ea-tamsd.final.2.real}), the TAMSD for the
non-equilibrium ensemble [case (B)]. However,
Eq.~(\ref{e.confine.ea-tamsd.eq}) is valid for arbitrary measurement times
$t$, while Eq.~(\ref{e.confine.ea-tamsd.final.2.real}) is valid only for
long measurement times ($1/\lambda \ll t$). In addition, the aging behavior
is absent in the equilibrium case.

\begin{figure}[t]
  \centerline{\includegraphics[width=8.2cm]{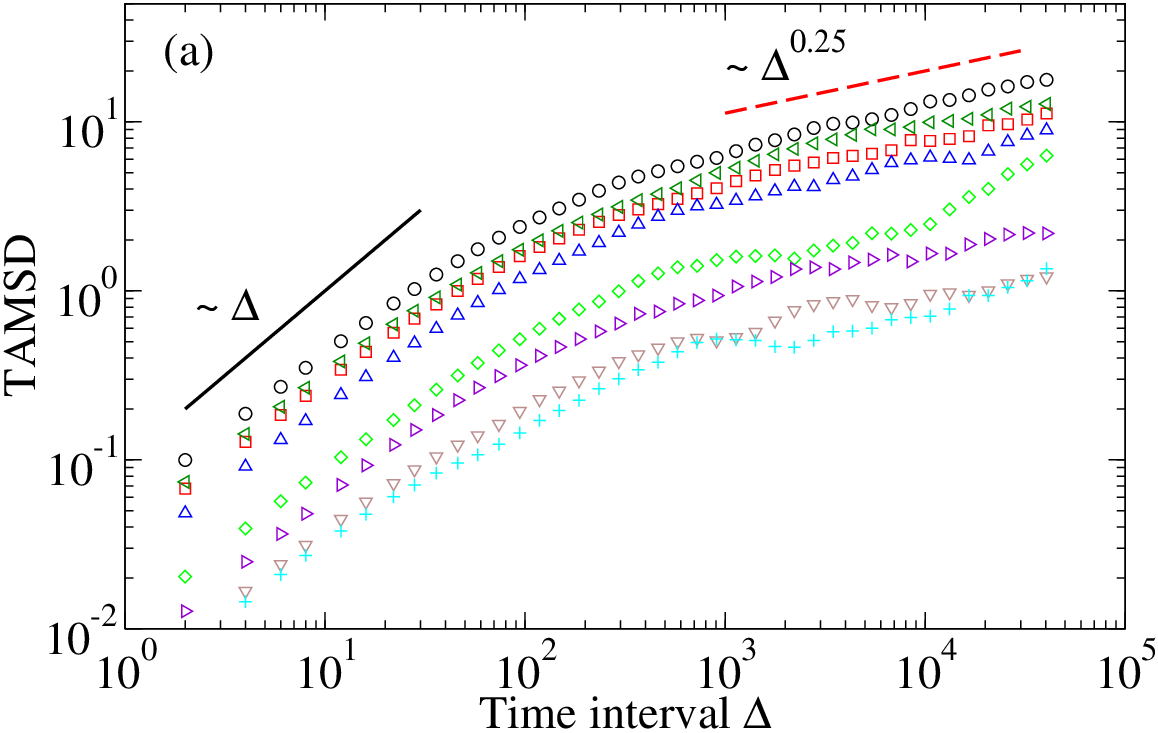}}
  \vspace*{.7cm}
  \centerline{\includegraphics[width=8.2cm]{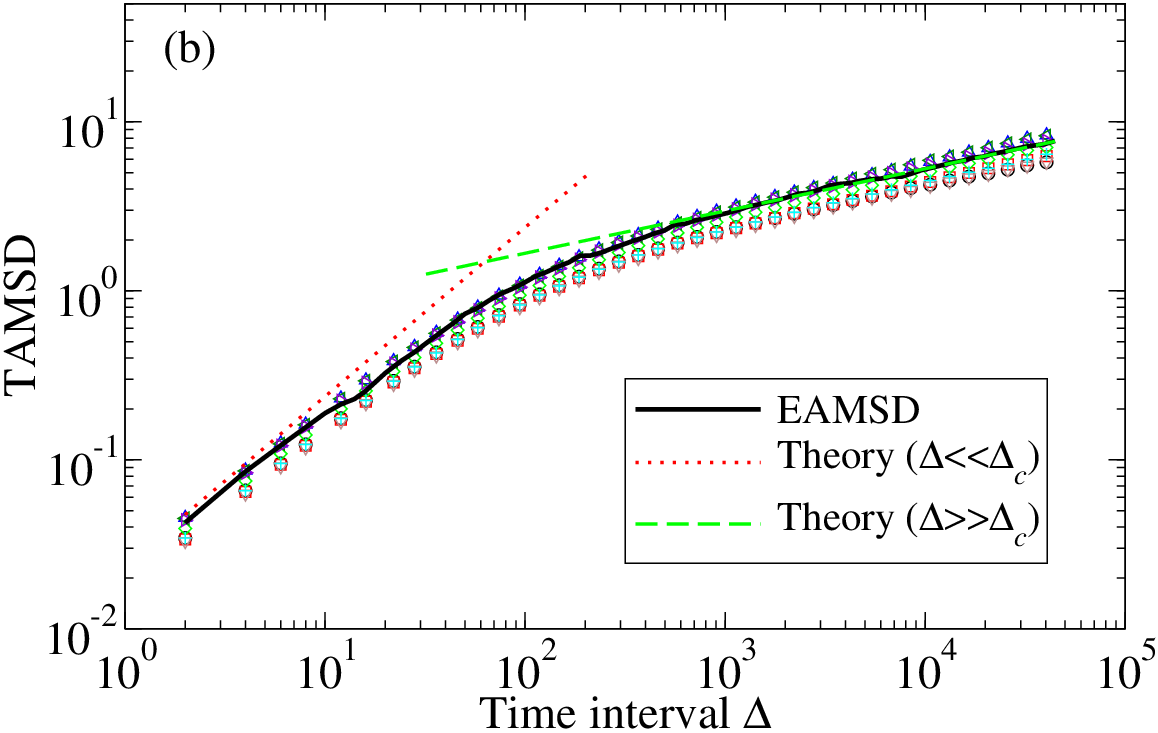}}
  \caption{\label{f.confine.msd.3} (Color online) (a) TAMSD
    $\overline{(\delta x)^2} (\Delta,t)$ vs. time interval $\Delta$ in
    log--log form for the equilibrium ensemble with $t<1/\lambda$. The
    total measurement time $t$ is set as $t=10^{6}$, and the cutoff
    parameter $\lambda$ as $\lambda=10^{-7}$. The other parameters are the
    same as in Fig.~\ref{f.confine.msd.1}. TAMSD is calculated for 8
    different realizations of trajectories and different symbols correspond
    to different realizations. (b) The same as the figure (a) except that
    $t=10^{8}$ ($t>1/\lambda$). The symbols are TAMSDs for 8 different
    trajectories and the thick solid curve is the EAMSD $\left\langle [
    x(\Delta)-x(0)]^2 \right\rangle_{\mathrm{eq}}$. The dotted and dashed
    lines are theoretical predictions given by
    Eq.~(\ref{e.confine.ea-tamsd.eq}). Note that no adjustable parameters
    were used to obtain these theoretical lines.  }
\end{figure}

In Fig.~\ref{f.confine.msd.3}(a) and (b), TAMSDs for 8 different
trajectories are shown for a short and long measurement times $t$,
respectively. Surprisingly, the scatter of TAMSD is even broader than that
in the non-equilibrium case [Fig.~\ref{f.confine.msd.1}(a)] at short
measurement times $t$ as shown in Fig.~\ref{f.confine.msd.3}(a).

In summary, for the non-equilibrium ensemble, the scatter of TAMSD appears
with the aging behavior [the non-equilibrium ensemble (case A)], whereas
for the equilibrium ensemble the scatter appears without aging. These
scatters are also studied quantitatively through theoretical analysis in
the next section. The theoretical predictions given by
Eq.~(\ref{e.confine.ea-tamsd.eq}) are also shown by dotted and dashed lines
in Fig.~\ref{f.confine.msd.3}(b).

Furthermore, from Eq.~(\ref{e.confine.ergode.eq}), the EAMSD for the
equilibrium initial ensemble $\left\langle [ x(\Delta)-x(0)]^2
\right\rangle_{\mathrm{eq}}$ is also given by the RHS of
Eq.~(\ref{e.confine.ea-tamsd.eq}). In Fig.~\ref{f.confine.msd.3}(b), a
numerically obtained EAMSD $\left\langle [ x(\Delta)-x(0)]^2
\right\rangle_{\mathrm{eq}}$ is displayed by a solid curve, which is
consistent with the theory [Eq.~(\ref{e.confine.ea-tamsd.eq}): the dotted
and dashed lines in Fig.~\ref{f.confine.msd.3}(b)].
Thus, the EAMSD for the equilibrium ensemble $ \left\langle [
x(\Delta)-x(0)]^2 \right\rangle_{\mathrm{eq}}$
[Eq.~(\ref{e.confine.ea-tamsd.eq})] is different from the EAMSD for the
non-equilibrium ensemble $ \left\langle [x(\Delta)-x(0)]^2 \right\rangle$
[Eq.~(\ref{e.confine.eamsd.5})].

\section {Statistical and ergodic properties of TAMSD \label{sec:STAT.TAMSD}}

The statistical property of TAMSD is dominated by the property of the
number of jumps $N_t$ until time $t$
(see Sec.~\ref{sec:stat.tamsd}). Therefore, we first study $N_t$ in
Secs.~\ref{s.real-space-analysis} and ~\ref{s.laplace-space-analysis} on
the basis of the analysis presented in \cite{miyaguchi11c}.

\subsection {Real space analysis \label{s.real-space-analysis}}

In the Sec.~\ref{sec:TSD}, we define $t_n$ as the time when the $n$-th jump
occurs for a trajectory $x(t)$: $t_n = \sum_{k=1}^{n} \tau_k$. From this
definition, we have the following relation:
\begin{eqnarray}
  G(n; t) &\equiv& \mathrm{Prob\,} (N_t < n)
  \nonumber\\[0.1cm]
  &=& \mathrm{Prob} (t_n > t)
  = \mathrm{Prob} \left( \sum_{k=1}^{n} \tau_k > t \right),
  \label{e.recipro.a}
\end{eqnarray}
where $\mathrm{Prob} (...)$ is the probability and $\tau_k~(k=1,2,...)$ are
the waiting times between jumps.

\subsubsection {Non-equilibrium ensemble}

From Eq.~(\ref{e.recipro.a}), we have the following equation for the
non-equilibrium ensemble:
\begin{eqnarray}
  \mathrm{Prob\,} (N_t < n) &=&
  \int_t^{\infty} d\tau ~P_{\rm TL}^{n} (\tau, \lambda)
  \nonumber\\[.1cm]
  &=&
  \int_{n^{-1/\alpha}t}^{\infty} d\tau ~P_{\rm TL} (\tau, n^{1/\alpha}\lambda),
  \label{e.recipro.b}
\end{eqnarray}
where we used the statistical independence between the waiting times
$\tau_k ~(k=1,2,...)$ and Eq.~(\ref{e.prob.na}). Furthermore, if we change
variables from $n$ to $x$ as $n = t^{\alpha} x$, we obtain
\begin{eqnarray}
  \label{e.cumulative.1}
  \mathrm{Prob} \left( \frac {N_t }{ t^{\alpha}} < x \right)
  &=&
  \int_{x^{-1/\alpha}}^{\infty} d\tau
  ~P_{\rm TL} \left(\tau, t x^{1/\alpha} \lambda \right)
  \nonumber\\[.1cm]
  &=&
  \int_{0}^{x} \frac {d\tau}{\alpha \tau^{1+1/\alpha}}
  ~P_{\rm TL} \left(\tau^{-1/\alpha}, t x^{1/\alpha} \lambda \right).
  \nonumber\\[-.2cm]
\end{eqnarray}
Note that the integrand of the RHS of Eq.~(\ref{e.cumulative.1}) is not the
PDF, because it contains the variable $x$. To derive the PDF for $x$, we
insert Eq.~~(\ref{e.prob.1b}) into Eq.~(\ref{e.cumulative.1}), and then we
have
\begin{eqnarray}
  \mathrm{Prob} \left( \frac {N_t }{ t^{\alpha}} < x \right)
  &=&
  -\frac {e^{c(t\lambda)^{\alpha}x}}{\alpha \pi}
  \sum_{k=1}^{\infty}
  \frac {\Gamma(k \alpha + 1)}{k!k}
  \nonumber \\[0.2cm]
  \label{e.cumulative.2}
  &&\times
  (-cx)^{k}
  \sin(\pi k \alpha) a_k,~~~~~
\end{eqnarray}
where
\begin{equation}
  \label{e.cumulative.a_k}
  a_k \equiv \int_{0}^{1} d\tau e^{-t \lambda \tau^{-1/(\alpha k)}}.
\end{equation}
Differentiating Eq.~(\ref{e.cumulative.2}) in terms of $x$, we obtain the
PDF for $x$:
\begin{eqnarray}
  f_{\lambda} (x; t)
  &=&
  -\frac {e^{c(t\lambda)^{\alpha}x}}{\alpha \pi}
  \sum_{k=1}^{\infty}
  \frac {\Gamma(k \alpha + 1)}{k!} (-c)^{k}
  \nonumber \\[0.2cm]
  \label{e.tml}
  &&\times
  \left[ \frac {c(t \lambda)^{\alpha} x}{k} + 1\right] x^{k-1}
  \sin(\pi k \alpha) a_k.~~~~~
\end{eqnarray}
When $\lambda = 0$, this PDF $f_{0}(x) \equiv f_{0}(x,t)$ is called the
Mittag--Leffler distribution \cite{aaronson97, akimoto10}.


\begin{figure}[t]
  \centerline{\includegraphics[width=8.2cm]{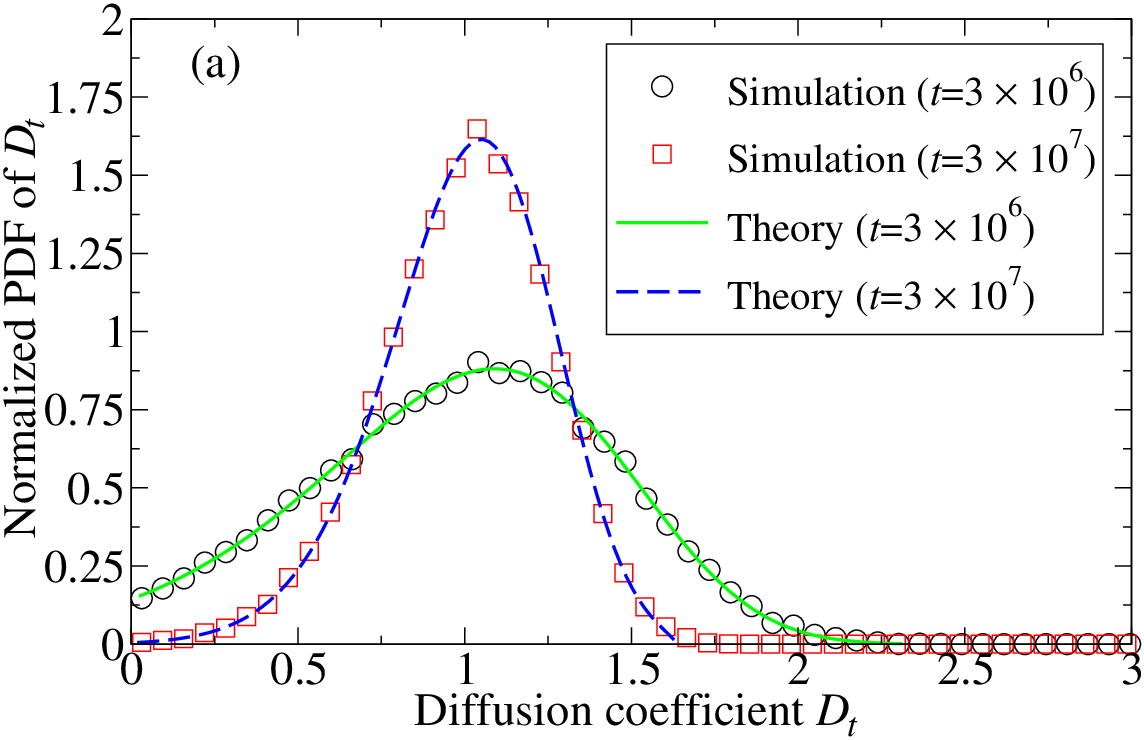}}
  \vspace*{.7cm}
  \centerline{\includegraphics[width=8.2cm]{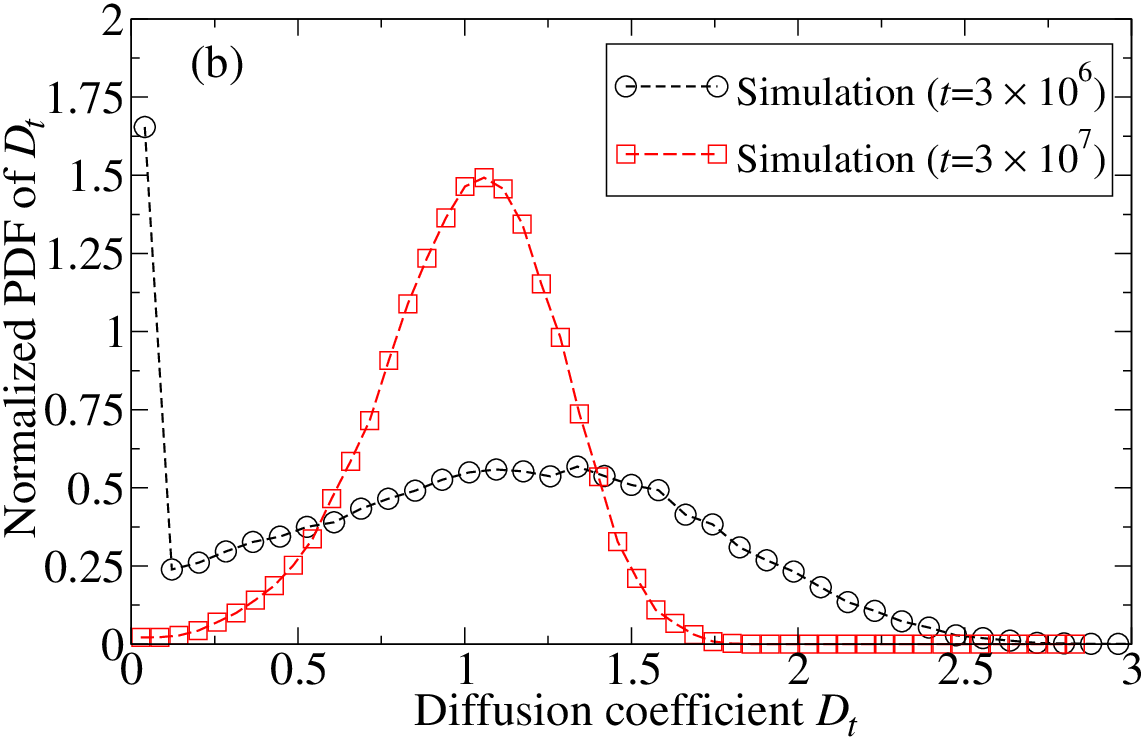}}
  \caption{\label{f.confine.ml} (color online) (a) PDF of the diffusion
    coefficient $D_t$ for the non-equilibrium TAMSD $ \overline{(\delta
      x)^2} (\Delta,t) \approx D_t \Delta $ for small $\Delta$. Each PDF is
    normalized so that its mean value equals unity. $D_t$ is calculated
    from TAMSD by least-square fitting over the interval $0 < \Delta <
    10$. The results for two different values of measurement times are
    presented: $t=3 \times 10^{6}$ (circles) and $3 \times 10^{7}$
    (squares). The other parameters are set as $\lambda=10^{-7},
    \alpha=0.75, c=1$, and $L=11$. The lines correspond to the theoretical
    predictions given by Eq.~(\ref{e.tml}). No adjustable parameters were
    used to obtain these curves. (b) PDF of the diffusion coefficient $D_t$
    for the equilibrium TAMSD $ \overline{(\delta x)^2} (\Delta,t) \approx
    D_t \Delta $ for small $\Delta$. The parameter values are the same as
    in the figure (a). The lines are to guide the eye.}
\end{figure}

As shown below [Sec.~\ref{sec:stat.tamsd}], TAMSD behaves as $
\overline{(\delta x)^2} (\Delta,t) \approx \frac {N_t}{t} \Delta $ for
small $\Delta$.  Thus, the diffusion constant $D_t$, defined as $
\overline{(\delta x)^2} (\Delta,t) \approx D_t \Delta $, shows the same
statistical property as that of $N_t$. Fig.~\ref{f.confine.ml} shows the
PDF of $D_t$ for two different measurement times $t$. Note that the PDF
narrows as $t$ increases. The analytical results given in Eq.~(\ref{e.tml})
are also depicted by solid and dashed lines in the figure.

\subsubsection {Equilibrium ensemble}

The equilibrium case is also analyzed in a similar way. In this case,
Eq.~(\ref{e.recipro.b}) should be replaced by
\begin{equation}
  \mathrm{Prob\,} (N_t < n) =
  \int_t^{\infty} d\tau \,
  \left(P_{\rm TL}^{\mathrm{eq}} \ast P_{\rm TL}^{n-1}\right) (\tau, \lambda),
  \label{e.recipro.b.eq}
\end{equation}
where $(f \ast g)$ means a convolution. In contrast to the non-equilibrium
case, however, it seems difficult to obtain a simple expression of the PDF
of $N_t$ for the equilibrium case. As shown in Fig.~\ref{f.confine.ml}, a
qualitative difference appears for short measurement time regime, $t <
1/\lambda$. In fact, there is a peak at $D_t = 0$ for the equilibrium
initial ensemble, which is due to the mean waiting time of $P_{\rm
  TL}^{\mathrm{eq}} (\tau, \lambda)$, $\left\langle \tau
\right\rangle_{\mathrm{eq}}$ ($\sim 1 /\lambda$), is much longer than that
of $P_{\rm TL} (\tau, \lambda)$, $\left\langle \tau \right\rangle$ ($\sim
1/\lambda^{1-\alpha}$). Namely, there are many trajectories that are
trapped more than the measurement time $t \,\,(< 1/\lambda)$
\cite{schulz12, *akimoto12}.

\subsection {Laplace space analysis \label{s.laplace-space-analysis}}
Next, to clarify the ergodic properties of the system, we study the
relative standard deviation (RSD) $R(t) \equiv \sqrt{\left\langle N_t^2
  \right\rangle_c} / \left\langle N_t \right\rangle$, where $\left\langle
\cdot \right\rangle_c$ is the cumulant. The quantity $R(t)$ is a measure of
ergodicity \cite{he08}. In fact, if $R(t) = 0$, time averages of an
observable give the same value independent of the trajectory; however, if
$R(t) > 0$, they are different from one trajectory to another. This
quantity $R(t)$ was also used in some molecular dynamics simulations to
characterize the ergodicity breaking and non-Gaussian fluctuations of lipid
motions in cell membranes \cite{akimoto11}, and to determine the longest
relaxation time in entangled polymers \cite{uneyama12}.

\subsubsection {Non-equilibrium ensemble}
Let us begin with the non-equilibrium ensemble. In order to derive an
analytical expression of $R(t)$, we start with the Laplace transformation
of Eq.~(\ref{e.recipro.b}):
\begin{eqnarray}
  \label{e.laplace.1}
  \tilde{G}(n; s)
  &=&
  \int_{0}^{\infty} dt e^{-ts}
  \int_{n^{-1/\alpha}t}^{\infty} d\tau ~P_{\rm TL} (\tau, n^{1/\alpha}\lambda)
  \nonumber\\[.2cm]
  &=&
  \frac {1 - e^{-nc[(\lambda + s)^{\alpha} - \lambda^{\alpha}]}}{s},
\end{eqnarray}
where $\tilde{G} (n; s)$ is the Laplace transformation of $G (n; t)$ in
terms of $t$. Next, let us define a function $g (n;s)$ as $g (n;s) \equiv
\tilde{G} (n+1; s) - \tilde{G} (n; s)$. Note that $g (n;s)$ is the Laplace
transformation (with respect to $t$) of the PDF of $N_t$. Furthermore, we
perform a discrete Laplace transformation of $g (n;s)$ with respect to $n$ as
follows:
\begin{eqnarray}
  \label{e.laplace.2}
  \tilde{g}(\nu; s) &=&
  \frac {1}{s}
  \frac
  {1- \exp \left( -c[(\lambda + s)^{\alpha} - \lambda^{\alpha}] \right)}
  {1- \exp \left( - \nu - c[(\lambda + s)^{\alpha} - \lambda^{\alpha}] \right)},
\end{eqnarray}
where we define $\tilde{g}(\nu; s)$ as $\tilde{g}(\nu; s) \equiv
\sum_{n=0}^{\infty} e^{-n\nu} g(n;s)$. Using the assumption $s, \lambda,
\nu \ll 1$, we have
\begin{eqnarray}
  \label{e.laplace.3}
  \tilde{g}(\nu; s)
  =
  \frac {1}{s}
  \sum_{k=0}^{\infty}
  \left(- \frac {\nu}{c}\right)^k
  \left[(\lambda + s)^{\alpha} - \lambda^{\alpha}\right]^{-k}.
\end{eqnarray}

\paragraph {First moments}
From Eq.~(\ref{e.laplace.3}), we obtain the Laplace transform
$\mathcal{L}[\left\langle N_t \right\rangle](s)$ of the first moment
$\left\langle N_t\right\rangle$ as
\begin{eqnarray}
  \label{e.laplace.1order.a}
  \mathcal{L}[\left\langle N_t \right\rangle](s)
  &\simeq&
  \left\{
  \begin{array}{ll}
    \displaystyle
    \frac {1}{c s^{\alpha+1}},& \hspace*{.1cm} s\gg \lambda \\[.3cm]
    \displaystyle
    \frac {1}{c \lambda^{\alpha-1} \alpha s^{2}}
    \left[ 1 + (1 - \alpha) \frac {s}{2 \lambda}\right],
    & \hspace*{.1cm} s \ll \lambda.~~~~~
  \end{array}
  \right.
\end{eqnarray}
The inverse Laplace transform of the above equation is given by
\begin{eqnarray}
  \label{e.laplace.1order.b}
  \left\langle N_t\right\rangle
  \simeq
  \left\{
  \begin{array}{ll}
    \displaystyle
    \frac {t^{\alpha}}{c \Gamma (\alpha + 1)},& \hspace*{.3cm} t \ll 1/\lambda \\[.3cm]
    \displaystyle
    \frac {t}{c \lambda^{\alpha-1} \alpha } + \frac {1 - \alpha}{2 c \lambda^{\alpha} \alpha },
    &\hspace*{.3cm} t \gg 1/\lambda.
  \end{array}
  \right.
\end{eqnarray}
The ensemble-averaged mean square displacement (EAMSD) for free CTRWs
(i.e., CTRWs without confinement effects) is known to be proportional to
$\left\langle N_t \right\rangle$ \cite{bouchaud90}: $\left\langle (\delta
\mbox{\boldmath $r$})^2 \right\rangle (t) \sim \left\langle N_t
\right\rangle$. Thus, the EAMSD of the present model without the effect of
confinement shows transient subdiffusion, i.e., subdiffusion for short
timescales and normal diffusion for long timescales \cite{saxton07}. The
crossover time between these two regimes is characterized by $1/\lambda$.

In addition, $\left\langle N_t\right\rangle$ is usually called the renewal
function in the renewal theory, and Eqs.~(\ref{e.laplace.1order.a}) and
(\ref{e.laplace.1order.b}) can also be derived by the renewal equation
\cite{cox62}. In contrast, higher order moments of $N_t$ cannot be derived
from the renewal equation, and we have to use Eq.~(\ref{e.laplace.3}).

\paragraph {Second moments}
Similarly, the Laplace transform $\mathcal{L}[\left\langle N_t^{2}
\right\rangle](s)$ of the second moment $\left\langle N_t^{2}\right\rangle$
is given by
\begin{eqnarray}
  \label{e.laplace.2order.a}
  \mathcal{L}[\left\langle N_t^{2} \right\rangle](s)
  &\simeq&
  \left\{
  \begin{array}{ll}
    \displaystyle
    \frac {2}{c^{2}s^{2\alpha+1}},& \hspace*{.1cm} s \gg \lambda \\[.3cm]
    \displaystyle
    \frac {2}{c^{2}\lambda^{2\alpha-2} \alpha^{2} s^{3}}
    \left[ 1 + (1 - \alpha) \frac {s}{\lambda}\right],
    & \hspace*{.1cm} s \ll \lambda.
  \end{array}
  \right.
  \nonumber\\[.0cm]
\end{eqnarray}
Then, the inverse transform is given by
\begin{eqnarray}
  \label{e.laplace.2order.b}
  \left\langle N_t^{2}\right\rangle
  \simeq
  \left\{
  \begin{array}{ll}
    \displaystyle
    \frac {2 t^{2 \alpha}}{c^{2}\Gamma (2 \alpha + 1)},& \hspace*{.3cm} t \ll 1/\lambda \\[.5cm]
    \displaystyle
    \frac {2}{c^{2} \lambda^{2\alpha-2} \alpha^{2} }
    \left[ \frac {t^{2}}{2} + \frac {1-\alpha}{\lambda}t\right],
    & \hspace*{.3cm} t \gg 1/\lambda.
  \end{array}
  \right.
\end{eqnarray}

\paragraph {Relative standard deviation}
Using Eqs.~(\ref{e.laplace.1order.b}) and (\ref{e.laplace.2order.b}), we
obtain asymptotics of the RSD $\sqrt{\left\langle N_t^2 \right\rangle_c} /
\left\langle N_t \right\rangle$:
\begin{eqnarray}
  \label{e.relative.sd}
  \frac {\sqrt{\left\langle N_t^2 \right\rangle_c}}{  \left\langle N_t \right\rangle}
  \simeq
  \left\{
  \begin{array}{ll}
    \displaystyle
    \sqrt{\frac {2 \Gamma^{2} (\alpha + 1)}{\Gamma (2 \alpha + 1)} - 1 },
    & \hspace*{.3cm} t \ll 1/\lambda \\[.5cm]
    \displaystyle
    \sqrt{
      \frac {1 - \alpha}{\lambda t}
    }
    & \hspace*{.3cm} t \gg 1/\lambda.
  \end{array}
  \right.
\end{eqnarray}
We define a crossover time $t_c$ between the two regimes, $t \ll 1/\lambda$
and $t \gg 1/\lambda$, as the intersection of the two functions in
Eqs.~(\ref{e.relative.sd}); therefore, we have
\begin{eqnarray}
  \label{e.tc}
  t_c &=&
  \frac {(1-\alpha)}
  {\frac {2 \Gamma^{2}(\alpha + 1)} {\Gamma (2\alpha + 1) }- 1}
  \lambda^{-1}.
\end{eqnarray}

\begin{figure}[tbh!]
  \centerline{\includegraphics[width=8.2cm]{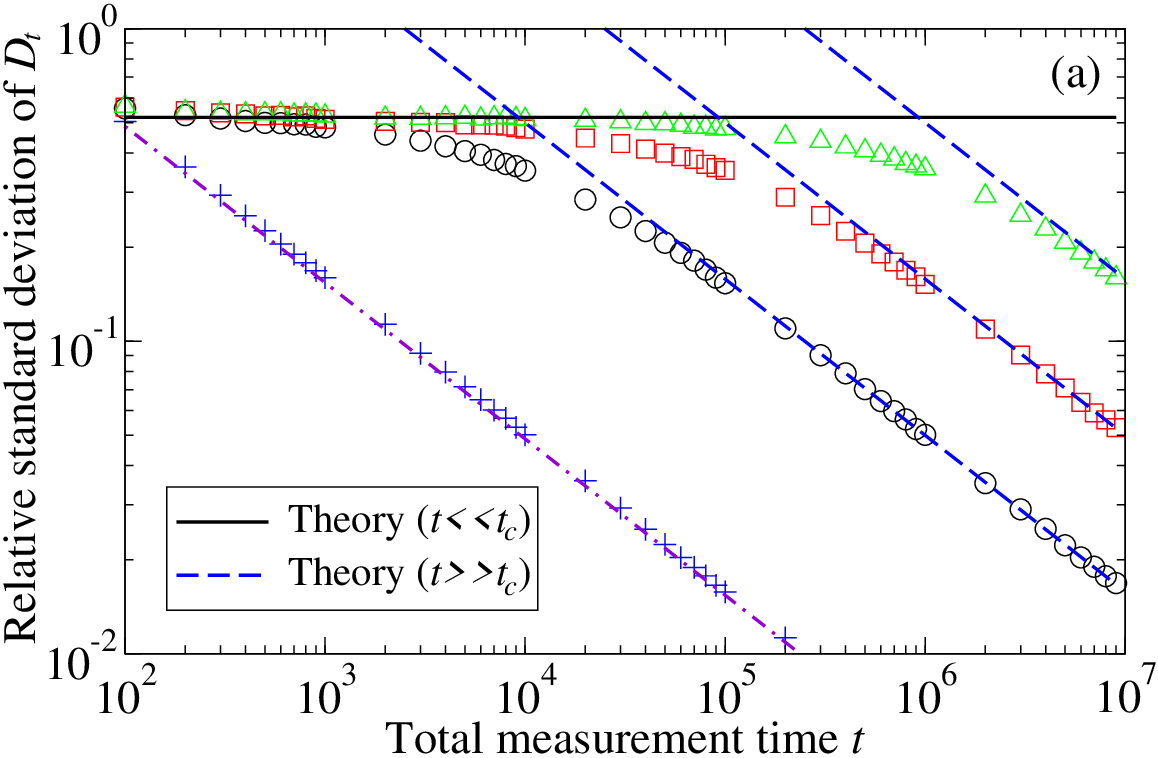}}
  \vspace*{0.7cm}
  \centerline{\includegraphics[width=8.2cm]{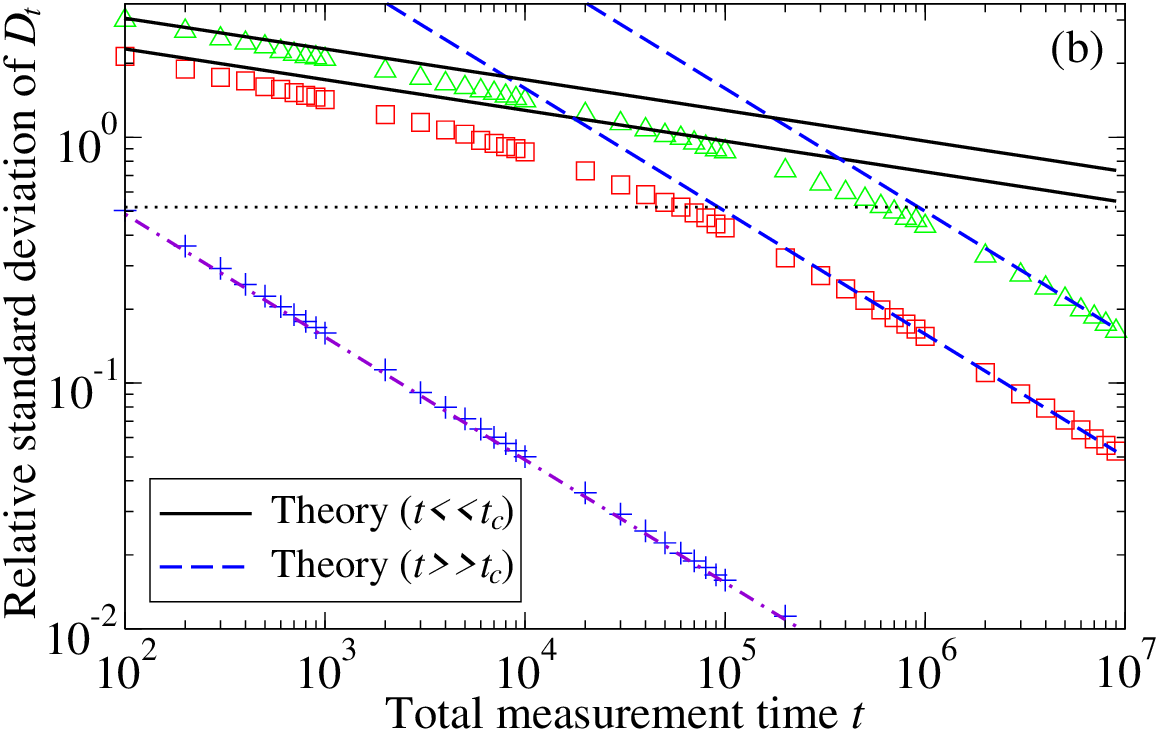}}
  \caption{\label{f.confine.rsd} (color online) (a) RSD $\sqrt{\left\langle
      D_t^2 \right\rangle_c} / \left\langle D_t \right\rangle$ vs. total
    measurement time $t$ for the non-equilibrium initial ensemble. $D_t$ is
    calculated from the TAMSD $\overline{(\delta x)^2} (\Delta,t)$ by a
    least-square fitting over the interval $0 < \Delta < 1$. Three
    different values of $\lambda$ are used: $\lambda=10^{-4}$ (circles),
    $10^{-5}$ (squares), and $10^{-6}$ (triangles). In addition, $\alpha$,
    $c$, and $L$ are set as $\alpha=0.75$, $c=1$, and $L = 11$,
    respectively. The lines correspond to the theoretical predictions given
    by Eq.~(\ref{e.relative.sd}); the solid line is the result for short
    timescales $t\ll t_c$, and the dashed lines are for long timescales $t
    \gg t_c$. The intersections of the solid and dashed lines correspond to
    the crossover times $t_c$ given by Eq.~(\ref{e.tc}). (b) RSD
    $\sqrt{\left\langle D_t^2 \right\rangle_{\mathrm{eq, }c}} /
    \left\langle D_t \right\rangle_{\mathrm{eq}}$ vs. total measurement
    time $t$ for the equilibrium initial ensemble. The parameter values are
    the same as those in the figure (a), except that the results only for
    two different values of $\lambda$ are plotted for clarity: $\lambda =
    10^{-5}$ (squares), and $10^{-6}$ (triangles). The lines correspond to
    the theoretical predictions given by Eq.~(\ref{e.relative.sd.eq}). (a)
    and (b) The plus signs in both figures are the RSD for the case in
    which the waiting time distribution is given by the exponential
    distribution $P(\tau) = \exp(\tau / \left\langle \tau \right\rangle) /
    \left\langle \tau \right\rangle$ with the same mean waiting time as
    that of the TSD with $\lambda = 10^{-6}$ (triangles): $\left\langle
    \tau \right\rangle = c \lambda^{\alpha-1} \alpha$. The dot--dashed line
    is a theoretical prediction for the exponential distribution:
    $R(t)=(c\alpha \lambda^{\alpha-1}/t)^{1/2}$. Note that the scales of
    vertical axes of two figures are different. For comparison, the
    theoretical prediction of RSD for the non-equilibrium ensemble at short
    timescales is depicted by a dotted line in the figure (b).}
\end{figure}

As shown in Fig.~\ref{f.confine.rsd}, the RSD remains almost constant
before the crossover time $t_c$, and starts decaying rapidly after the
crossover.  In the figure, the RSD for the exponential waiting-time
distribution, which has the same mean waiting time $\left\langle \tau
\right\rangle$ as that of the TSD with $\lambda = 10^{-6}$, is also shown
by pluses. It is clear that the RSD for the exponential distribution
(pluses) decays much more rapidly than that for the TSD (triangles).


\subsubsection {Equilibrium ensemble}

The calculation for the equilibrium case is almost parallel to that for the
non-equilibrium case except that we should use Eq.~(\ref{e.recipro.b.eq})
instead of Eq.~(\ref{e.recipro.b}). Thus, we only show the final results.
First, the generating function $\tilde{g}(\nu; s)$ is given by
\begin{equation}
  \label{e.laplace.4}
  \tilde{g}(\nu; s)
  =
  \frac {1}{\lambda^{\alpha -1} \alpha s^2}
  \sum_{k=0}^{\infty}
  \left(- \frac {\nu}{c}\right)^k
  \left[(\lambda + s)^{\alpha} - \lambda^{\alpha}\right]^{-k+1}.
\end{equation}
Using this function, we obtain $\left\langle N(t)
\right\rangle_{\mathrm{eq}}$
\begin{equation}
  \left\langle N_t \right\rangle_{\mathrm{eq}}
  \simeq \frac {t}{c\lambda^{\alpha-1}\alpha},
  \label{e.mean.nt.eq}
\end{equation}
and the RSD for the equilibrium initial ensemble
\begin{eqnarray}
  \label{e.relative.sd.eq}
  \frac {\sqrt{\left\langle N_t^2 \right\rangle_{\mathrm{eq},c}}}
  {\left\langle N_t \right\rangle_{\mathrm{eq}}}
  \simeq
  \left\{
  \begin{array}{ll}
    \displaystyle
    \sqrt{\frac {2 \alpha }{\Gamma (2 + \alpha) (\lambda t)^{1-\alpha}}},
    & \hspace*{.3cm} t \ll 1/\lambda \\[.5cm]
    \displaystyle
    \sqrt{
      \frac {1 - \alpha}{\lambda t}
    }
    & \hspace*{.3cm} t \gg 1/\lambda.
  \end{array}
  \right.
\end{eqnarray}
Although the RSD slowly decays at the short measurement timescales, the
relative fluctuations are even larger than the non-equilibrium case as
shown in Fig.~\ref{f.confine.rsd}.

\subsection {Statistical properties of TAMSD \label{sec:stat.tamsd}}

In this section, we discuss the statistical properties of TAMSD for
one-dimensional CTRWs, although the analysis for free CTRWs in the
followings can be generalized for higher dimensional systems
\cite{miyaguchi11c}. Properties applicable to a general class of
observables are summarized in Appendix~\ref{sec:hopf}.

Here, we show that the TAMSD can be approximately given by the time average
of the following observable:
\begin{eqnarray}
  \label{e.hopf.f(t')}
  h(t')
  &=&
  \sum_{k=1}^{\infty} \delta(t'- t_k) h_k
  \\[0.cm]
  \label{e.hopf.Hk}
  h_k &=& \Delta z_k^2 + 2 \sum_{l=1}^{k-1} z_k z_l \theta (\Delta - (t_k - t_l)),
\end{eqnarray}
where $z_k = \pm 1$ is the displacement of the jump at time $t'=t_k$, and
$\theta (t)$ is a step function defined by
\begin{eqnarray}
  \label{e.hopf.theta(t)}
  \theta (t) =
  \left\{
  \begin{array}{ll}
    0,~~~&(t<0)
    \\[0.2cm]
    t,~~~&(t\geq 0).
  \end{array}
  \right.
\end{eqnarray}
Let us express a trajectory of a CTRW as
\begin{equation}
  \label{e.hopf.x(t')}
  x(t') = \sum_{k=1}^{\infty} z_k I (t_k < t'),
\end{equation}
where $I (t_k < t')$ is the indicator function defined as follows: $I (t_k
< t') = 1$ if the inside of the bracket is satisfied, while $I (t_k < t') =
0$ otherwise.  Then, the displacement $x(t'+\Delta) - x(t')$ is given by
\begin{equation}
  \label{e.hopf.dx(t')}
  x(t'+ \Delta) - x(t')
  =
  \sum_{k=1}^{\infty} z_k I (t' < t_k < t' + \Delta).
\end{equation}
Furthermore, the squared displacement $[x(t'+ \Delta) - x(t')]^2$ is
expressed as
\begin{eqnarray}
  [x(t'+ \Delta) - x(t')]^2
  &=& \sum_{k=1}^{\infty} I (t_k - \Delta < t' < t_k) z_k^2
  \nonumber\\[0.0cm]
  &&
  + 2 \sum_{k=1}^{\infty}
  \sum_{l=1}^{k-1}
  z_k z_l I (t_k-\Delta < t' < t_l).
  \nonumber\\[0.cm]
  \label{e.hopf.x(t').sq}
\end{eqnarray}
From Eqs.~(\ref{e.hopf.x(t').sq}) and (\ref{e.tamsd.1}), we obtain the
following approximation for the TAMSD:
\begin{equation}
  \overline{(\delta x)^2} (\Delta,t)
  \approx
  \frac {1}{t}
  \sum_{k=1}^{N_t} \left[
  \Delta z_k^2 + 2  \sum_{l=1}^{k-1} z_k z_l \theta (\Delta - (t_k - t_l))
  \right].
  \label{e.hopf.tamsd}
\end{equation}
Now, it is clear that the RHS of Eq.~(\ref{e.hopf.tamsd}) is equivalent to
the time average of $h(t')$, defined by Eqs.~(\ref{e.hopf.f(t')}) and
(\ref{e.hopf.Hk}).

\subsubsection {CTRWs without reflecting boundaries}

In the absence of the confinement effect \cite{miyaguchi11c} (i.e., without
reflecting boundaries), the relations $\left\langle h_k \right\rangle =
\Delta$ and $\left\langle h_k h_{k+n} \right\rangle - \left\langle h_k
\right\rangle \left\langle h_{k+n} \right\rangle = 0$ for $n \geq 1$ hold
because of the mutual independence of $z_k$, $\left\langle z_k
\right\rangle = 0$ and $ z_k^2 = 1$. In addition, since $z_k$ is
independent of the initial ensemble, the same relations can be obtain also
for the equilibrium ensemble: $\left\langle h_k \right\rangle_{\mathrm{eq}}
= \Delta$ and $\left\langle h_k h_{k+n} \right\rangle_{\mathrm{eq}} -
\left\langle h_k \right\rangle_{\mathrm{eq}} \left\langle h_{k+n}
\right\rangle_{\mathrm{eq}} = 0$.  Then, $h_k$ [Eq.~(\ref{e.hopf.Hk})]
satisfies the law of large numbers [Eq~.(\ref{e.app.hopf.2})]; therefore,
we have
\begin{equation}
  \label{e.hopf.noconfine}
  \overline{(\delta x)^2} (\Delta,t)
  \approx
  \frac {1}{t} \sum_{k=1}^{N_t} h_k
  =
  \frac {N_t}{t} \frac {1}{N_t} \sum_{k=1}^{N_t} h_k
  =
  \frac {N_t}{t} \Delta,
\end{equation}
for arbitrary $\Delta$. Note that this result is independent of the choice
of the initial ensembles.  Therefore, TAMSD increases in proportion to
$\Delta$, and has the same statistical property as $N_t$. In particular,
the diffusion constant $D_t$ is given by
\begin{eqnarray}
  \label{e.hopf.d.const}
  D_t = \frac {N_t}{t}.
\end{eqnarray}
Furthermore, the PDF of $D_t$ is given by Eq.~(\ref{e.tml}), and the RSD of
$D_t$ shows the asymptotics such as the ones given in
Eq.~(\ref{e.relative.sd}).

\subsubsection {CTRWs with reflecting boundaries}

Although a similar property holds for the system with reflecting
boundaries, the calculation becomes more complicated, because
$z_k~(k=1,2,...)$ are not mutually independent. Here, we present only an
outline of the proof. The main tool is the spectral decomposition of the
$n$-time transition matrix $P^{(n)}$ of the DTRWs with reflecting
boundaries \cite{feller68c16}:
\begin{equation}
  \label{e.decay.corr.tmatrix}
  P^{(n)} =
  \sum_{r=0}^{L-1}
  \left| r \right\rangle \left\langle r \right| \lambda_r^n,
\end{equation}
where $L$ is the total number of sites.  The ket $\left| r \right\rangle$
is the $L$-dimensional eigenvector of the transition matrix defined by
\begin{equation}
  \left| 0 \right\rangle =
  \begin{pmatrix}
    \frac {1}{L^{1/2}}\\
    \vdots\\
    \frac {1}{L^{1/2}}
  \end{pmatrix},
  \quad
  \label{e.decay.corr.eigen.vec.1}
\end{equation}
for $r=0$, and
\begin{equation}
  \left| r \right\rangle =
  \begin{pmatrix}
    r_1\\
    \vdots\\
    r_L
  \end{pmatrix}
  \quad
  \text{with}
  \quad
  r_j =
  \frac { \sin \frac {\pi r j}{L} - \sin \frac {\pi r (j-1)}{L}}
  {L^{1/2} (1 - \cos \frac {\pi r}{L})^{1/2}},
  \label{e.decay.corr.eigen.vec}
\end{equation}
for $1 \leq r \leq L - 1$. Here, $j$ is a site index (1 $\leq j \leq L$).
Also, the bra $\left\langle r \right|$ is defined by the transpose of
$\left| r \right\rangle$. The real numbers $\lambda_r$ are the associated
eigenvalues: $\lambda_0 = 1$ for $r=0$ and $\lambda_r = \cos (\pi r/L)$ for
$1 \leq r \leq L-1$. Thus, the first term ($r=0$) in
Eq.~(\ref{e.decay.corr.tmatrix}) corresponds to the eigenmode with the unit
eigenvalue (the non-decaying mode), which is a uniform state, and the other
terms decay exponentially fast to zero. Each element of the transition
matrix, i.e., $P_{i,j}^{(n)}$, is the transition probability from the site
$i$ to the site $j$ during $n$ jumps. An important point is that each
element of the matrix tends to $1/L$ exponentially fast as $n$
increases. It follows that correlation functions for the DTRWs decay
exponentially, too. By using this fact, we can show (after some lengthy
calculations) that the correlation functions for the CTRWs $\left\langle
h_k h_{k+n} \right\rangle - \left\langle h_k \right\rangle \left\langle
h_{k+n} \right\rangle$, which can be expressed with correlation functions
for the DTRWs such as $ \left\langle z_nz_lz_kz_1 \right\rangle -
\left\langle z_nz_l\right\rangle \left\langle z_k z_1 \right\rangle$, also
decay exponentially.

As a result, the law of large numbers holds for the $h_k$ even for the
confined system, and we obtain
\begin{equation}
  \label{e.hopf.confine}
  \overline{(\delta x)^2} (\Delta,t)
  \approx
  \frac {N_t}{t} \frac {1}{N_t} \sum_{k=1}^{N_t} h_k
  =
  \frac {N_t}{t} \mu_h \left(\Delta \right).
\end{equation}
The function $\mu_h (\Delta)$ can be derived by using the transition matrix
[Eq.~(\ref{e.decay.corr.tmatrix})] with a suitable hydrodynamic limit,
however an easier way here is to utilize the results in
Sec.~\ref{sec:EA.TAMSD}. For example, from
Eqs.~(\ref{e.confine.ea-tamsd.final.1.real}) and (\ref{e.laplace.1order.b})
[or Eqs.~(\ref{e.confine.ea-tamsd.final.2.real}) and
(\ref{e.laplace.1order.b})] we obtain

\begin{align}
  \label{e.tamsd.final}
  \overline{(\delta x)^2} (\Delta,t)
  \approx
  \begin{cases}
    \dfrac {N_t}{t} \Delta, & \text{for} \quad \Delta \ll \Delta_c
    \\[0.3cm]
    \dfrac {N_t}{t} \dfrac {cL^2\Delta^{1 - \alpha}}{6\Gamma (2 - \alpha)},
    & \text{for} \quad \Delta \gg \Delta_c,
  \end{cases}
\end{align}
for the non-equilibrium initial ensemble. Similarly, from
Eqs.~(\ref{e.confine.ea-tamsd.eq}) and (\ref{e.mean.nt.eq}), we have
Eq.~(\ref{e.tamsd.final}) also for the equilibrium ensemble. Thus,
Eq.~(\ref{e.tamsd.final}) is valid for both non-equilibrium and equilibrium
ensembles. Thus, all the differences between the two ensembles are included
in the statistical properties of $N_t$ in Eq.~(\ref{e.tamsd.final}). See,
for example, Eqs.~(\ref{e.laplace.1order.b}) and (\ref{e.mean.nt.eq}).

\section {Conclusion \label{sec:conclusion}}

Up to now, two characteristic properties have been reported for confined
CTRWs \cite{neusius09}: (i) there is a crossover from normal to anomalous
diffusion, and (ii) TAMSDs are distributed depending on trajectories (i.e.,
weak ergodicity breaking). These results are for the non-equilibrium
ensemble [There is no equilibrium ensemble for CTRWs with power-law waiting
times with stable index $\alpha \in (0, 1)$, because the system never
reaches an equilibrium state.]. In this paper, in addition to this
confinement effect, we incorporated a cutoff into the waiting time
distribution. For the case of the non-equilibrium ensemble, we analytically
determined that the property (i) persists even in the long measurement time
limit $t \to \infty$ (This fact was first found numerically in
\cite{burov11}). In contrast, the property (ii), that is, the
distributional behavior of TAMSDs, appears for short measurement times $t$,
whereas ergodicity holds for longer measurement times. The important point
is that as compared to common distributions such as exponential
distribution, it takes a very long time to observe ergodic behavior for the
case where the waiting time distribution is given by the TSD. In addition,
this transition from weak ergodicity breaking to ergodicity is a transition
from an irreproducible regime to a reproducible regime.

Furthermore, as shown in Eq.~(\ref{e.tc}), the crossover time $t_c$ between
weak ergodicity breaking and ergodicity is proportional to $1/\lambda$
[also, the crossover time between anomalous and normal diffusion in EAMSD
for free CTRWs is proportional to $1/\lambda$; see the text below
Eq.~(\ref{e.laplace.1order.b})]. Because the mean waiting time
$\left\langle \tau \right\rangle$ is given by $\left\langle \tau
\right\rangle \sim 1/\lambda^{1-\alpha}$, we have
\begin{eqnarray}
  \label{e.tc.tau-average}
  t_c \sim \left\langle \tau \right\rangle^{\frac {1}{1-\alpha}}.
\end{eqnarray}
Thus, the crossover time $t_c$ is not proportional to the mean waiting time
$\left\langle \tau \right\rangle$; in fact, $t_c$ can be much longer than
$\left\langle \tau \right\rangle$ for $\alpha$ close to 1. These facts may
be important for estimating the crossover time in experiments \cite{xu11}.

In contrast to the CTRWs with power-law waiting times with stable index
$\alpha \in (0, 1)$, the CTRWs with cutoff waiting times have an
equilibrium state. It is surprising that the crossover from normal to
anomalous diffusion [property (i)], and a scatter in TAMSD [property (ii)]
exist even for the case of the equilibrium initial ensemble. The scatter in
TAMSD is even broader than the case of the non-equilibrium ensemble.  The
main difference from the non-equilibrium case is that there is no aging for
the equilibrium ensemble. Another difference between the two ensembles is
that the RSD decays algebraically even in the short time regime for the
equilibrium case, whereas it does not decay for the non-equilibrium
case. These properties for the equilibrium ensemble might fit well with
some experimental results \cite{golding06, bronstein09, *kepten11, burov11,
  jeon11} (See the next section). The important point is that the absence
of aging in experimental data does not necessarily exclude the possibility
of the CTRWs. We also presented a numerical method to generate the
equilibrium ensemble (Appendix \ref{sec:num.method}).

\section {Discussion \label{sec:discussion}}

First, we compare the model studied in this paper (in particular the
equilibrium CTRWs) with the experimental data presented in
\cite{jeon11}. In \cite{jeon11}, they studied lipid granules in a harmonic
potential, and used confined non-equilibrium CTRWs as a theoretical
model. They presented the following quantitative agreements between the
experimental data and the model: (1) the crossover from normal to anomalous
diffusion in TAMSD and (2) the scatter distribution of the TAMSD. They set
the total measurement time $t = 3$ sec for these data, which seems to
correspond to the case (B) in our classification
[Eq.~(\ref{e.confine.timescales.2}): namely, a time regime after
aging]. See the Fig.~6 in the supplementary material of \cite{jeon11}.

However, there seems to be one disagreement between the experimental data
and the non-equilibrium CTRWs: i.e., (3) the aging property. In fact, the
non-equilibrium CTRWs show aging in short measurement times [time regime
(A) in Eq.~(\ref{e.confine.timescales.2})], whereas the experimental data
do not (see the Fig.~6 in the supplementary material).

On the other hand, the equilibrium CTRWs reproduce all three properties. In
the time regime (B) [Eq.~(\ref{e.confine.timescales.2})], the equilibrium
CTRWs show quantitatively the same behavior as the non-equilibrium CTRWs
with respect to the properties (1) and (2). Thus, the equilibrium CTRWs
will also reproduce these experimental observations successfully. In
addition, the equilibrium CTRWs show no aging even in the time regime (A),
and thus the model agrees the experimental data with respect to the
property (3), too.

Second, although we show that time averages for a class of observables
including TAMSD follow the ML distribution (Appendix \ref{sec:hopf}),
different distributions can arise for different classes of observables
\cite{rebenshtok07, *rebenshtok08, *akimoto08b, *saa10}. As an example, we
show in Appendix \ref{sec:hopf} a case in which long time averages follow
the PDF called the bilateral ML distribution \cite{kasahara77}.

Third, there are qualitatively different types of random walk models in
random environments. Recently, anomalous properties have been found in both
experiments \cite{harada09, golding06, weber10, weber10a, bronstein09,
  xu11, banks05, jeon11, burov11} and molecular dynamics simulations
\cite{neusius11, akimoto11, *uneyama12}. The CTRWs, which are a model of
random walks in random environments, are often used as a model of these
anomalous properties. However, there are similar random walk models in
random environments with qualitatively different statistical behavior: for
example, barrier models \cite{alexander81, bouchaud90}, random force models
\cite{bouchaud90, slutsky04a, *slutsky04b} and the reptation model for
entangled polymers \cite{doi78, uneyama12}. To the best of the authors'
knowledge, the ergodic properties, such as the behavior of RSD, of these
systems are still unclear.

Finally, besides random walks in random environments, there are still
several different mechanisms for anomalous behavior, e.g., GLE, FBMs, and
diffusion on fractal structures \cite{havlin02}. Therefore, it is important
to develop techniques for time series analysis to elucidate which mechanism
is the actual cause of the anomalous properties observed in various
experiments and molecular dynamical simulations \cite{magdziarz10,
  magdziarz11b}. It is also important to investigate systems in which these
mechanisms are combined \cite{meroz10}, and those in which nonlinear
dynamics are coupled with these anomalous mechanisms \cite{goychuk09,
  harada09}.



\begin{acknowledgments} We thank S.~Shinkai for drawing our attention to
  Ref.\cite{kasahara77}, and T.~Uneyama for fruitful discussions concerning
  possible connections between CTRWs and entangled polymer systems.  In
  addition, we are grateful to an anonymous referee for his helpful
  comments on Eq.~(\ref{e.confine.frec_time}) and Appendix
  \ref{sec:app.eq10}. This study is partially supported by a Grant-in-Aid
  for Young Scientists (B) (22740262).
\end{acknowledgments}
\appendix {}
\section {Infinitely divisible distribution \label{sec:app.TSD}}

Here we briefly introduce the one-sided TSD. In general, the infinitely
divisible distributions $P(\tau)$ is defined as follows: if $\omega (s)$ is
the characteristic function (Fourier transform) of $P(\tau)$, then there
exists a characteristic function $\omega_n(s)$ such that $\omega (s) =
\omega_n^n (s)$. Note that this is not trivial, because the function
$\omega^{1/n} (s)$ is not necessarily a characteristic function of a PDF
(i.e., a non-negative function).

Now let us define the one-sided TSD $P_{\rm TL} (\tau, \lambda)$. First, we
define the characteristic function $e^{\psi (\zeta, \lambda)}$ of $P_{\rm
  TL} (\tau, \lambda)$ as follows:
\begin{eqnarray}
  \label{e.prob.1a}
  P_{\rm TL}(\tau, \lambda) &=&
  \frac {1}{2 \pi}
  \int_{-\infty}^{\infty}
  e^{\psi (\zeta, \lambda)} e^{- i \zeta \tau} d\zeta,
  \\[0.2cm]
  e^{\psi (\zeta, \lambda)}
  &=&
  \int_{-\infty}^{\infty} P_{\rm TL}(\tau, \lambda) e^{i\zeta \tau} d\tau.
\end{eqnarray}
Then, the function $\psi (s, \lambda)$ is defined by the canonical form of
the infinitely divisible distributions %
\cite{feller71}:
\begin{eqnarray}
  \label{e.cfunc1}
  \psi (\zeta, \lambda)
  =
  \int_{-\infty}^{\infty}
  \left(
  e^{i\zeta \tau} - 1
  \right)
  f (\tau, \lambda) d\tau.
\end{eqnarray}
If $f(\tau, \lambda)$ is a PDF in terms of $\tau$, then $e^{\psi (\zeta,
  \lambda)}$ is the characteristic function of the compound Poisson
distribution \cite{feller71}. For the TSD, however, the function $f (\tau,
\lambda)$ is not a PDF, and is defined as follows \cite{koponen95,
  magdziarz11a, nakao00, cartea07, *del-Castillo-Negrete09,
  *stanislavsky09, *stanislavsky11, gajda10}:
\begin{eqnarray}
  \label{e.incriment}
  f (\tau, \lambda)
  =
  \left\{
  \begin{array}{ll}
    0, &  ~~~(\tau<0)
    \\[0.1cm]
    \displaystyle
    - c \, \frac {\tau^{-1-\alpha} e^{-\lambda \tau}}{\Gamma(-\alpha)}, &  ~~~(\tau>0),
  \end{array}
  \right.
\end{eqnarray}
where $c$ is the scale factor, and $\alpha$ and $\lambda$ are parameters
satisfying $0 < \alpha < 1$ and $\lambda \geq 0$, respectively. By changing
the integration path on the complex plane in Eq.~(\ref{e.cfunc1}) to $C: z
= s(\lambda + i\zeta)/(\lambda^2 + \zeta^2)~[s\in [0, \infty)]$, we obtain
$\psi (\zeta, \lambda)$ as
\begin{eqnarray}
  \label{e.cfunc3}
  \psi (\zeta, \lambda)
  =
  -c \left[ (\lambda - i \zeta)^{\alpha} - \lambda^{\alpha} \right].
\end{eqnarray}
The above equation indicates an important property:
\begin{eqnarray}
  \label{e.infinite.divide}
  n \psi (\zeta, \lambda) = \psi (n^{1/\alpha} \zeta, n^{1/\alpha}\lambda),
\end{eqnarray}
which is related to the infinite divisibility introduced in the first
paragraph of this section, because it can be rewritten as $\psi
(n^{-1/\alpha} \zeta, n^{-1/\alpha} \lambda) = \psi(\zeta, \lambda) /n$,
and $\psi (n^{-1/\alpha} \zeta, n^{-1/\alpha} \lambda)$ is obviously a
characteristic function. It is clear that Eq.~(\ref{e.prob.na}) is obtained
from Eq.~(\ref{e.infinite.divide}).


Now, we derive Eq.~(\ref{e.prob.1b}). From Eqs.~(\ref{e.prob.1a}) and
(\ref{e.cfunc3}), we have
\begin{eqnarray}
  \label{e.prob.1c}
  P_{\rm TL} (\tau, \lambda) &=&
  \frac {e^{c \lambda^{\alpha}}}{2 \pi}
  {\rm Re\,}
  \int_{0}^{\infty} e^{-c \left( \lambda - i \zeta \right)^{\alpha}}
  e^{-i \zeta \tau} d \zeta
  \nonumber\\
  &=&
  \frac {e^{c \lambda^{\alpha}}}{ \pi \tau}
  {\rm Im\,}
  \int_{0}^{\infty} e^{-c \left( \lambda - \frac {s}{\tau} \right)^{\alpha}}
  e^{-s} d s,
\end{eqnarray}
where we have changed the integration path from the real to the imaginary
axis by using the Cauchy's integral theorem.  Then, we have
\begin{eqnarray}
  \label{e.prob.1d}
  P_{\rm TL} (\tau, \lambda) &=&
  \frac {e^{c \lambda^{\alpha} - \lambda \tau}}{ \pi \tau}
  {\rm Im\,}
  \int_{-\tau \lambda}^{\infty} e^{-c \left( -\frac {s}{\tau} \right)^{\alpha}}
  e^{-s} d s
  \nonumber\\
  &=&
  \frac {e^{c \lambda^{\alpha} - \lambda \tau}}{ \pi \tau}
  {\rm Im\,}
  \int_{0}^{\infty} e^{-c \left( -\frac {s}{\tau} \right)^{\alpha}}
  e^{-s} d s.
\end{eqnarray}
Finally, we obtain Eq.~(\ref{e.prob.1b}) by Taylor expansion of the
exponential function $e^{-c \left( -\frac {s}{\tau} \right)^{\alpha}}$ and
the integral representation of the Gamma function $\Gamma (k\alpha+1) =
\int_0^{\infty} s^{\alpha k}e^{-s}ds$.


The equilibrium waiting time distribution (\ref{e.prob.1b.eq}) can be
derived by using Eq.~(\ref{e.app.rv.equil}) as follows:
\begin{equation}
  P_{\rm TL}^{\mathrm{eq}} (\tau, \lambda) =
  \int_{0}^{1} \frac {\tau}{a \left\langle \tau \right\rangle}
  P_{\rm TL} \left(\frac {\tau}{a}, \lambda\right) da.
\end{equation}

\section {Method of Numerical Simulation \label{sec:num.method}}

Let $Y_{\lambda}$ be a random variable following the TSD
[Eq.~(\ref{e.prob.1b})]. We briefly review the method to generate
$Y_{\lambda}$ numerically according to \cite{gajda10}. First, a random
variable $Y_{0}$, which follows the one-sided stable distributions, can be
generated by the following equation \cite{chambers76, *weron96,
  *janicki93}:
\begin{eqnarray}
  \label{e.rv_levy}
  Y_{0}
  &=&
  c^{ 1/ \alpha}
  \frac
  {\sin \left(\alpha \left(V + \frac {\pi}{2} \right) \right)}
  {\left( \cos V\right)^{1/\alpha}}
  \nonumber
  \\
  &&
  \times
  \left[
  \frac {\cos \left(V - \alpha \left(V + \frac {\pi}{2} \right)\right)}{W}
  \right]^{(1-\alpha)/\alpha},
\end{eqnarray}
where $V$ is a uniform noise in $[-\pi/2, \pi/2]$, and $W$ is an exponential
noise with mean 1.  To generate $Y_{\lambda}$, note that TSD
[Eq.~(\ref{e.prob.1b})] is just a stable distribution multiplied by the
exponential factor $e^{- \lambda \tau}$. Therefore, we first generate
$Y_{0}$, then we accept it with a probability $\exp(- \lambda Y_0) < 1$. If
$Y_{0}$ is rejected, it will be regenerated according to
Eq.~(\ref{e.rv_levy}) until it is accepted. Thereafter, the random variable
that is finally accepted follows TSD \cite{gajda10}.

Similarly, to generate equilibrium noise $Y_{\lambda}^{eq}$, first we
create an auxiliary random variable $T$ by accepting $Y_0$ with a
probability $\lambda Y_0\exp(- \lambda Y_0 + 1)$. Then, $Y_{\lambda}^{eq}$
is obtained by $Y_{\lambda}^{eq} = XT$, where $X$ is a uniform noise in
$[0,1]$. This can be proved as follows.

First, let us consider a time axis $(-\infty, \infty)$ covered by many
non-overlapping time intervals. Each interval is obtained from
$P_{\mathrm{TL}} (\tau, \lambda)$ (The intervals are assumed to be mutually
independent). Then, choose a time $t$ randomly from the time axis. Now, the
random variable $T$ is defined by the length of the time interval which
includes the randomly chosen time $t$. In addition, let $\rho (\tau)$ be
the PDF of the variable $T$. Then, $\rho (\tau)$ is given by
\begin{equation}
  \rho (\tau) = \frac {\tau } {\left\langle \tau \right\rangle}
  P_{\mathrm{TL}} (\tau, \lambda)
  \propto
  \tau e^{-\lambda \tau} P_{\mathrm{TL}} (\tau, 0),
\end{equation}
because the probability with which an interval is chosen is proportional to
the length of that interval $\tau$. Thus, the random variable $T$ can be
generated from the stable noise $Y_{0}$ by accepting it with the
probability $\lambda Y_0\exp(- \lambda Y_0 + 1) \leq 1$ (The factor
$\lambda e$ is added to make the maximum probability be unity for
computational efficiency.). Moreover, by the Laplace transform of the above
equation, we have
\begin{equation}
  \label{e.app.rv.auxiliary}
  \tilde{\rho} (s) = - \frac {1}{\left\langle \tau \right\rangle}
  \frac {\partial \tilde{P}_{\mathrm{TL}}}{\partial s} (s, \lambda).
\end{equation}

Next, let us define a new variable $Y_{\lambda}^{\mathrm{eq}} = XT$, where
$X$ is the uniform noise in $[0,1]$. Then, we have
\begin{equation}
  \mathrm{Prob} \left(Y_{\lambda}^{\mathrm{eq}} > \tau \right)
  =
  \mathrm{Prob} \left( XT > \tau \right)
  =
  \int_{0}^{1} \mathrm{Prob} \left( T > \frac {\tau}{a} \right) da.
\end{equation}
By differentiating the both side, we obtain
\begin{equation}
  \label{e.app.rv.equil}
  \rho_{\mathrm{eq}} (\tau)
  =
  \int_{0}^{1} \rho \left(\frac {\tau}{a}\right) \frac {da}{a},
\end{equation}
where $\rho_{\mathrm{eq}} (\tau)$ is the PDF of
$Y_{\lambda}^{\mathrm{eq}}$. The Laplace transform of the above equation
leads to
\begin{equation}
  \tilde{\rho}_{\mathrm{eq}} (s) = \int_{0}^{1} \tilde{\rho} (as) da.
\end{equation}
Inserting Eq.~(\ref{e.app.rv.auxiliary}) into the above equation, we obtain
$\tilde{\rho}_{\mathrm{eq}} (s) = [1 - \tilde{P}_{\mathrm{TL}} (s,
\lambda)]/\left\langle \tau \right\rangle s$, which is the Laplace
transform of the equilibrium waiting time PDF
[Eq.~(\ref{e.laplace.of.ptl.eq})]. Therefore, $Y_{\lambda}^{\mathrm{eq}}$
follows the equilibrium waiting time distribution
$P_{\mathrm{TL}}^{\mathrm{eq}} (\tau, \lambda)$.

This method to generate the equilibrium waiting time might be applicable to
other probability densities with finite moments.

\section {Time translation invariance of the equilibrium waiting time \label{sec:app.time.trans}}

Let $w(\tau)$ be a waiting time distribution for a renewal process, and
$w^{\mathrm{eq}}(\tau)$ be the associated equilibrium waiting time
distribution.  Here, we show the relation
$w_{\mathrm{e}}^{\mathrm{eq}}(\tau; t') = w^{\mathrm{eq}}(\tau)$
[Eq.~(\ref{e.time_translation_inv})] by the method presented in
\cite{godrche01}.  First, let us define
\begin{equation}
  w(\tau; t', n) \equiv
  \left\langle
  \delta \left(\tau - (t_{n+1} - t')\right)
  I \left( t_n < t' < t_{n+1} \right)
  \right\rangle,
\end{equation}
where $I(a < t' < b) = 1$ if the inside of the bracket is satisfied, 0
otherwise.  In addition, $t_n\,(n=1, 2, \dots)$ are renewal times defined
by $t_n \equiv \sum_{k=1}^{n} \tau_k$ , where $\tau_k$ are successive
waiting times.  The Laplace transform with respect to $\tau$ and $t'$ gives

\begin{align}
  \breve{w} (u; s, n) &=
  \left\langle
  \int_{t_n}^{t_{n+1}} e^{-(t_{n+1} - t') u} e^{-t's} dt'
  \right\rangle
  \notag\\[0.1cm]
  &=
  \begin{cases}
    \dfrac {\tilde{w}^{\mathrm{eq}} (u) - \tilde{w}^{\mathrm{eq}} (s)}{s-u},
    &\text{for} \,\, n=0
    \\[0.1cm]
    \tilde{w}^{\mathrm{eq}}(s)\tilde{w}^{n-1} (s) 
    \dfrac {\tilde{w} (u) - \tilde{w} (s)} {s-u},
    &\text{for} \,\, n\geq 1.
  \end{cases}
\end{align}
Because $\tilde{w} (u; s) =  \sum_{n=0}^{\infty}\tilde{w} (u; s, n)$, we have
\begin{equation}
  \label{e.app.inv.w_e.eq}
  \breve{w}_{\mathrm{e}}^{\mathrm{eq}}(u; s) =
  \frac {\tilde{w} (u) - \tilde{w} (s)} {s-u}
  \cdot
  \frac {\tilde{w}^{\mathrm{eq}}(s)}{1 - \tilde{w} (s)}
  +
  \frac {\tilde{w}^{\mathrm{eq}} (u) - \tilde{w}^{\mathrm{eq}} (s)}{s-u}.
\end{equation}
Here, if we replace $\tilde{w}^{\mathrm{eq}}(s)$ with $\tilde{w}(s)$, we
obtain the forward recurrence time distribution for the non-equilibrium
ensemble [Eq.~(\ref{e.confine.w_e})]:
\begin{equation}
  \breve{w}_{\mathrm{e}}(u; s) =
  \frac {\tilde{w} (u) - \tilde{w} (s)} {s-u}
  \cdot
  \frac 1{1 - \tilde{w} (s)}.
\end{equation}
It follows that
\begin{equation}
  \label{e.app.inv.w.eq}
  \tilde{w}^{eq}(u) = \lim_{s \to 0} s
  \tilde{w}_{\mathrm{e}}(u; s) =
  \frac {1 - \tilde{w} (u)}{\left\langle \tau \right\rangle u}.
\end{equation}
By inserting Eq.~(\ref{e.app.inv.w.eq}) into Eq.~(\ref{e.app.inv.w_e.eq}),
we find that $\breve{w}_{\mathrm{e}}^{\mathrm{eq}}(u; s) =
\tilde{w}^{eq}(u) /s$. The inverse Laplace transform gives the relation we
want.

\section {Derivation of Eq.~(\ref{e.confine.frec_time}) \label{sec:app.eq10}}

In this appendix, we present a detailed derivation of
Eq.~(\ref{e.confine.frec_time}), which we owe to an anonymous referee.

Here, we consider the non-equilibrium initial ensemble, i.e., $t=0$ is
the renewal time (the particle jumps at $t=0$). First, we write the LHS
of Eq.~(\ref{e.confine.frec_time}) as
\begin{equation}
  \label{e.app.eq10.1}
  \left\langle [ x(t'+\Delta)-x(t')]^2 \right\rangle
  =
  \int_{-\infty}^{\infty}dx x^2 l(x;t', t'+\Delta),
\end{equation}
where $l(x; t', t'+\Delta)$ is the PDF of the displacement in the interval
$t\in [t', t'+\Delta]$. Furthermore, $l(x; t', t'+\Delta)$ can be rewritten
as
\begin{equation}
  \label{e.app.eq10.2}
  l(x; t', t'+\Delta) = \sum_{n=0}^{\infty} P (n;t', t'+\Delta) l^{\ast n} (x),
\end{equation}
where $P (n;t', t'+\Delta)$ is the probability of having $n$ jumps in the
interval $t\in [t', t'+\Delta]$, $l(x)$ is the PDF of single jump length,
and $l^{\ast n} (x)$ is the $n$-times convoluted PDF of $l(x)$ with
$l^{\ast 0} (x) = \delta(x)$.  Note that
we set $l(x) = [\delta (x+1) + \delta (x-1)]/2$ in our model.
Moreover, the following recursion relation is the most essential ingredient:
\begin{align}
  \label{e.app.eq10.3}
  P (n; t', t'+\Delta) &=
  \int_{0}^{\Delta}d\tau w_e(\tau;t') P (n-1; t'+\tau, t'+\Delta)\notag\\[0.1cm]
  &=
  \int_{0}^{\Delta}d\tau w_e(\tau;t') P (n-1; 0, \Delta-\tau)
\end{align}
for $n\geq1$. Note that we can replace $t'+\tau$ with $0$ (and $t'+\Delta$
with $\Delta - \tau$), because both $t'+\tau$ and $0$ are the renewal
times.

Putting these equations together, we obtain
\begin{align}
  &\left\langle [ x(t'+\Delta)-x(t')]^2 \right\rangle\notag\\[0.2cm]
  &=
  \int_{0}^{\Delta}d\tau w_e(\tau;t')
  \sum_{n=0}^{\infty} P (n; 0, \Delta-\tau)
  \int_{-\infty}^{\infty} dx x^2 l^{\ast (n+1)} (x).
\end{align}
Now, if we put $l(x) = [\delta (x+1) + \delta (x-1)]/2$ into the
convolution of the above equation, $l^{\ast (n+1)} (x) = [l^{\ast n} \ast
l](x)$, we have Eq.~(\ref{e.confine.frec_time}).  
\section {Derivation of GFFPE \label{sec:appc}}
GFFPE was derived from a subordinated process in \cite{gajda10}. Here, we
derive GFFPE from CTRWs in a hydrodynamic limit \cite{montroll65,
  bouchaud90, metzler00}. Let $P(x,t)$ be the PDF of a particle at a time
$t$, and $P_0(x)$ be the initial density $P_0(x) \equiv P(x,0)$. (In this
section, we consider $t$ to be the usual time variable, instead of the
total measurement time).  Moreover, each particle is assumed to follow the
CTRW dynamics on the real line $x$ or a one-dimensional lattice $x= x_0,
x_{\pm 1},x_{\pm 2},...$. Now, we define $\psi (x, t) dxdt$ as the
probability of a particle to perform a jump with length $x$ after being
trapped for a certain period $t$. Then, the probability of a particle to be
trapped for a period $t$ is given by
\begin{eqnarray}
  \label{e.appc.phi}
  \phi (t)
  =
  1 - \int_{-\infty}^{\infty}dx' \int_{0}^{t} \psi (x', t') dt'
  =
  1 - \int_{0}^{t} w (t') dt'.\qquad
\end{eqnarray}
Furthermore, we define $Q(x,t) dt dx$ as the probability of a particle to
reach an interval $[x,x+dx)$ in the period $[t, t+dt)$. Then, we have
\begin{eqnarray}
  \label{e.appc.defs}
  P (x,t) &=& \int_{0}^{t} dt' \phi (t-t') Q(x,t') + \phi(t) P_0(x),
  \\[0.2cm]
  Q (x,t) &=&
  \int_{-\infty}^{\infty} dx' \int_{0}^{t} dt' \psi (x', t') Q (x-x', t-t')
  \nonumber\\[0.1cm]
  && + \int_{-\infty}^{\infty} dx' \psi (x', t) P_0(x-x')
\end{eqnarray}
Taking the Fourier and Laplace transforms with respect to space and time,
respectively, we obtain
\begin{eqnarray}
  \label{e.appc.fourier-laplace.2}
  \tilde{Q} (k,u)
  &=&
  \frac {\tilde{\psi} (k, u) \tilde{P}_0(k)}{1 - \tilde{\psi} (k,u)},
  \\[0.2cm]
  \label{e.appc.fourier-laplace.3}
  \tilde{P} (k,u)
  &=&
  \frac {1 - \tilde{w}(u)}{u}
  \frac {\tilde{P}_0(k)}{1 - \tilde{\psi} (k,u)},~~~
\end{eqnarray}
where we used the relation $\tilde{\phi}(u) = (1 - \tilde{w}(u))/u$ [the
Laplace transform of Eq.~(\ref{e.appc.phi})].

Next, we assume that the PDF $\psi (x,t)$ can be separable, i.e., $\psi (x,
t) = l (x) w(t)$.
We further assume that
\begin{eqnarray}
  \label{e.appc.lambda}
  \tilde{l} (k)&\simeq& 1 - \frac {\left\langle \delta x^{2}\right\rangle}{2}k^{2}
\end{eqnarray}
in the hydrodynamic limit $k \to 0$, where $\left\langle \delta
x^{2}\right\rangle$ is the mean squared displacement of a single jump,
i.e.,~$\left\langle \delta x^{2}\right\rangle = \int_{-\infty}^{\infty} dx
~x^{2}l(x)$.  For one-dimensional CTRWs with jumps only to the nearest
neighbor sites, we have $\left\langle \delta x^{2}\right\rangle= 1$.
Moreover, we use the TSD, $P_{\rm TL} (t)$, as the waiting time
distribution $w(t)$:
\begin{eqnarray}
  \label{e.appc.tlf}
  \tilde{w}(u)
  &\simeq&
  1 - c \left[ (\lambda + u)^{\alpha} - \lambda^{\alpha} \right],
\end{eqnarray}
where $\lambda, u \ll 1$ is assumed. Under these assumptions,
Eq.~(\ref{e.appc.fourier-laplace.3}) can be rewritten as follows:
\begin{eqnarray}
  \label{e.appc.ffpe.1}
  u\tilde{P} (k,u) - \tilde{P}_0(k)
  \simeq -
  \frac {\left\langle \delta x^{2}\right\rangle}{2c}
  \frac {u k^2} {\left[ (\lambda + u)^{\alpha} - \lambda^{\alpha} \right]}
  \tilde{P} (k,u).\qquad
\end{eqnarray}
The inverse Fourier and Laplace transformations with respect to space and
time lead to
\begin{eqnarray}
  \label{e.appc.ffpe.3}
  \frac {\partial P(x,t)}{\partial t}
  =
  K \hat{\Phi}_t  \frac {\partial^2 P (x,t)}{\partial x^2},
\end{eqnarray}
where $K$ is a constant given by $K= {\left\langle \delta x
  \right\rangle^{2}} / {2c}$, and $\hat{\Phi}_t$ is an operator defined by
\begin{eqnarray}
  \label{e.appc.ffpe.4}
  \hat{\Phi}_t f(t) &\equiv& \frac {d}{dt} \int_{0}^{t} M(t-t') f(t') dt',
\end{eqnarray}
with function $M(t)$, which is defined by its Laplace transform:
\begin{eqnarray}
  \label{e.appc.ffpe.5}
  \tilde{M} (u)
  =
  \int_{0}^{\infty} dt e^{-ut} M(t) dt
  =
  \frac {1}{(\lambda + u)^{\alpha} - \lambda^{\alpha}}.
\end{eqnarray}
As $\lambda \to 0$, Eq.~(\ref{e.appc.ffpe.3}) leads to the usual FFPE
\cite{neusius09, metzler00}. Thus, the above equations
(\ref{e.appc.ffpe.3})--~(\ref{e.appc.ffpe.5}) are called generalized FFPE
\cite{gajda11}. Note also that GFFPE is the equation for non-equilibrium
initial ensemble. In contrast, in order to describe the system started with
the equilibrium ensemble, the multi-point fractal diffusion equation would
be necessary
\cite{baule07,*barkai07,*politi11,*meerschaert12,*meerschaert12b}.

\section {Mixed problem of GFFPE \label{sec:appd}}
The initial-boundary value problem of GFFPE [Eq.~(\ref{e.appc.ffpe.3})]
under the boundary condition
\begin{eqnarray}
  \label{e.appd.mix.2}
  \frac {\partial P}{\partial x} (0, t) = \frac {\partial P}{\partial x} (L, t) = 0,
\end{eqnarray}
and initial condition
\begin{eqnarray}
  \label{e.appd.mix.3}
  P(x,0) = \delta (x - x_s)
\end{eqnarray}
can be solved by a standard method for the diffusion equation
\cite{neusius09}.
In fact, by assuming the separability $P(x,t) = X(x)T(t)$, we obtain
\begin{eqnarray}
  \label{e.appd.separable.2}
  \frac {T'(t)}{K \hat{\Phi}_t T(t)} = \frac {X''(x)}{X(x)} = -q,
\end{eqnarray}
where $q$ is a constant. $X(x)$ satisfying the boundary condition
[Eq.~(\ref{e.appd.mix.2})] is given by $X_n(x) = \cos (n\pi x/L)$ and $q =
\left( {n\pi}/{L} \right)^{2}$~($n=0,1,2...$).  Similarly, $T(t)$ is
derived in its Laplace form:
\begin{eqnarray}
  \label{e.appd.separable.t.laplace.2}
  \tilde{T_n}(u) =
  \frac {T_n(0)}{u + (n\pi / L)^2 K u \tilde{M}(u)}.
\end{eqnarray}
Then, we have the solution in the following form: ${P}(x,t) =
\sum_{n=0}^{\infty} X_n(x) {T}_n(t)$. Finally, $T_n(0)$ in
Eq.~(\ref{e.appd.separable.t.laplace.2}) can be determined from the initial
condition [Eq.~(\ref{e.appd.mix.3})] as
\begin{eqnarray}
  \label{e.appd.initial.2}
  T_0(0) = \frac {1}{L},\quad
  T_n(0) = \frac {2}{L} \cos  \frac {n \pi  x_s}{L}~~(n=1,2,...).\qquad
\end{eqnarray}
Finally, we obtain
\begin{eqnarray}
  \label{e.appd.mixed}
  P(x,u)
  =
  \frac {1}{Lu}
  +
  \frac {2}{L}
  \sum_{n=1}^{\infty}
  \frac {
    \cos \frac {n \pi x_s}{L}
    \cos \frac {n \pi x  }{L}  }
  {u + (n\pi / L)^2 K u \tilde{M}(u)}. \quad
\end{eqnarray}
Note that $P(x,u)$ in the above equation is just the transition probability
$P(x,u;x_s,0)$ used in Sec.~\ref{sec:EA.TAMSD}.

\section {Riemann zeta function \label{sec:appe-riemann}}
First, we define the functions $\zeta^{e} (\beta)$ and $\zeta^{o} (\beta)$ as
\begin{eqnarray}
  \label{e.appe.zeta}
  \zeta^{e} (\beta) \equiv
  \sum_{\begin{subarray}{c}n=2 \\ n: \mathrm{even}\end{subarray} }^{\infty}
  \frac {1}{n^{\beta}},
  \hspace*{0.7cm}
  \zeta^{o} (\beta) \equiv
  \sum_{\begin{subarray}{c}n=1 \\ n: \mathrm{odd}\end{subarray}}^{\infty}
  \frac {1}{n^{\beta}}.
\end{eqnarray}
By using the integral representation of the gamma function,
\begin{eqnarray}
  \label{e.appe.power.int}
  \frac {1}{n^{\beta}}
  =
  \frac {1}{\Gamma(\beta)}
  \int_{0}^{\infty} dv v^{\beta-1} e^{-nv},
\end{eqnarray}
it is easy to obtain the following formula:
\begin{eqnarray}
  \label{e.appe.even.zeta.int}
  \zeta^{e} (\beta) &=&
  \frac {\zeta(\beta)}{2^{\beta}},
\end{eqnarray}
where $\zeta(\beta)$ is the Riemann zeta function. From
Eq.~(\ref{e.appe.even.zeta.int}), we have
\begin{eqnarray}
  \label{e.appe.odd.zeta}
  \zeta^{o} (\beta) = \frac {2^{\beta}-1}{2^{\beta}} \zeta(\beta).
\end{eqnarray}
In particular, we have $\zeta^{o} (4) = \pi^{4} / 96$ and $\zeta^{o} (2) =
\pi^{2} / 8$.

\section {Ergodic theorems for general observables \label{sec:hopf}}
In this appendix, we summarize the ergodic properties of general
observables. As an example, we derive the spatial distribution of free
CTRWs.
First, we consider a renewal process with renewal times $t_k~(k=1,2,\dots)$
(see Sec.~\ref{sec:TSD}). We also define observables that take nonzero
values only at the renewal times $t'=t_k~(k=1,2,\dots)$:
\begin{eqnarray}
  \label{e.app.hopf.1}
  h(t') = \sum_{k=1}^{\infty} H_k \delta(t'-t_k),
\end{eqnarray}
where $\{H_k\}$ are random variables with the same mean value $\left\langle
H_k \right\rangle = \mu_h$ ($k=1,2,...$). $\{H_k\}$ are assumed to be
independent of $N_t$, but not necessarily mutually independent. We also
assume the ergodicity with respect to the operational time, i.e., the
number of renewals $n$:
\begin{eqnarray}
  \label{e.app.hopf.2}
  \frac {1}{n}\sum_{k=1}^{n} H_k
  \simeq
  \mu_h,~~~{\rm as}~~~n \to \infty.
\end{eqnarray}
This relation is just the law of large numbers. A sufficient condition for
Eq.~(\ref{e.app.hopf.2}) is that the correlation function $C(n) \equiv
\left\langle H_k H_{k+n} \right\rangle - \left\langle H_k \right\rangle
\left\langle H_{k+n} \right\rangle$ decays faster than $n^{-\gamma}~(\gamma
> 0)$ \cite{bouchaud90, burov10}.

From Eq. (\ref{e.app.hopf.1}), the time average of the function $h(t')$
is given by
\begin{eqnarray}
  \label{e.app.hopf.3}
  \frac {1}{t} \int_{0}^{t} dt' h (t')
  =
  \frac {1}{t} \sum_{k=1}^{N_t} H_k.
\end{eqnarray}
Note that the value of the RHS of Eq.~(\ref{e.app.hopf.3}) depends on the
trajectories of the renewal process in general. For example, if $H_k\equiv
1 ~(k=1,2,...)$ and the PDF of the renewal time $\tau$ is give by a power
law $w(\tau) \sim 1/\tau^{1+\alpha}$ with $0 <\alpha < 1$, the RHS of the
above equation is equivalent to $N_t/t$, which follows the ML distribution
[Eq.~(\ref{e.tml}) with $\lambda = 0$] as $t \to \infty$.

First, we consider the case in which $\mu_h \neq 0$. In this case, we
obtain the following equation from Eq.~(\ref{e.app.hopf.2}):
\begin{eqnarray}
  \label{e.app.hopf.4}
  \frac {1}{N_t}\sum_{k=1}^{N_t} H_k \to \mu_h,
  \quad \text{as} \quad t \to \infty.
\end{eqnarray}
Thus, Eq.~(\ref{e.app.hopf.3}) can be rewritten as
\begin{eqnarray}
  \label{e.app.hopf.5}
  \frac {1}{t} \int_{0}^{t} dt' h (t')
  \simeq
  \frac {N_t}{t} \mu_h
\end{eqnarray}
Similarly, if another observable $g(t')$ defined by $g (t')=
\sum_{k=1}^{\infty} G_k \delta (t' -t_k)$ satisfies the same conditions as
$h(t')$, we have
\begin{eqnarray}
  \label{e.app.hopf.f.2}
  \frac {\int_{0}^{t} g(t') dt'} {\int_{0}^{t} h(t') dt'}
  \to
  \frac {\mu_g}{\mu_h},
  \quad \text{as} \quad t \to \infty.
\end{eqnarray}
Note that the RHS is not a random variable. Equation (\ref{e.app.hopf.f.2})
is a stochastic version of Hopf's ergodic theorem for dynamical systems
\cite{aaronson97}.

Next, we consider the case $\mu_h = 0$. We assume that the correlation
function $C(n)$ decays faster than $1/n$; then, from the central limit
theorem, we have
\begin{eqnarray}
  \label{e.app.hopf.6}
  \frac {1}{n^{1/2}}\sum_{k=1}^{n} H_k
  \simeq
  F_{\sigma},~~~{\rm as}~~~n \to \infty,
\end{eqnarray}
where $F_{\sigma}$ is a random variable following the Gaussian distribution
with mean 0 and variance $\sigma^2$ \cite{bouchaud90, ma85}. The variance
$\sigma^2$ is given by $\sigma^{2} = C(0) + 2 \sum_{n=1}^{\infty} C(n)$
\cite{bouchaud90}. From Eqs.~(\ref{e.app.hopf.3}) and (\ref{e.app.hopf.6}),
we obtain
\begin{eqnarray}
  \label{e.app.hopf.7}
  \frac {1}{t} \int_{0}^{t} dt' h (t')
  \simeq
  \frac {N_t^{1/2}}{t} F_{\sigma}.
\end{eqnarray}
If we define a random variable $Y_t$ as $Y_t \equiv (N_t/t^{\alpha})^{1/2}
F_{\sigma}$, its characteristic function $\left\langle e^{i\xi
  Y_t}\right\rangle$ is given by
\begin{equation}
  \label{e.app.hopf.8}
  \left\langle e^{i\xi Y_t}\right\rangle
  =
  \tilde{f}_{\lambda}
  \left(
  \frac {\sigma^2 \xi^2}{2}, t
  \right),
\end{equation}
where $\tilde{f}_{\lambda} (\xi, t)$ is the Laplace transform of the PDF
$f_{\lambda} (x, t)$ [Eq.~(\ref{e.tml})]. Although, in general, it is
difficult to obtain the PDF of $Y_t$, we can derive it explicitly for
$\lambda = 0$. In this case, since $\tilde{f}_{0} (\xi) = \tilde{f}_{0}
(\xi, t)$, which is the Laplace transform of the ML distribution, we obtain
\begin{equation}
  \label{e.app.hopf.9}
  \tilde{f}_{0} (\xi)
  =
  \sum_{k=0}^{\infty} \left( - \frac {\xi}{c} \right)^{k}
  \frac {1}{\Gamma (\alpha k +1)}.
\end{equation}
To derive this equation, let us introduce an auxiliary variable $h$ in
Eq.~(\ref{e.cumulative.1}) as $\mathrm{Prob} \left( N_t/ t^{\alpha} < xh
\right) = \int_{hx^{-1/\alpha}}^{\infty} d\tau ~P_{\rm TL} \left(\tau, 0
\right)$. Then double Laplace transformations with respect to $h$ and $x$
give $\frac {1}{s\xi} \sum_{k=0}^{\infty} \left(-
\xi/cs^{\alpha}\right)^{k}$, where $s$ is the Laplace variable conjugate to
$h$. Finally, taking the inverse Laplace transform with respect to $h$ and
setting $h = 1$, we obtain Eq.~(\ref{e.app.hopf.9}). Here note that
$\mathrm{Prob} \left( N_t / t^{\alpha} < xh \right) $ is the integral of
the PDF $f_{0} (x,t)$. By using Eqs.~(\ref{e.app.hopf.8}) and
(\ref{e.app.hopf.9}), we obtain the characteristic function for $\lambda=0$
explicitly:
\begin{equation}
  \label{e.app.hopf.10}
  \left\langle e^{i\xi Y_t}\right\rangle
  =
  \sum_{k=0}^{\infty} \left( \frac {i\xi}{\sqrt{2c}/\sigma} \right)^{2k}
  \frac {1}{\Gamma ((\alpha/2) \cdot 2k +1)}.
\end{equation}
Note that the PDF of $Y_t$ is symmetric with respect to the $y$-axis from
its definition and that the even order moments of $Y_t$ are equivalent to
those of the ML distribution with index $\alpha/2$ and scale factor
$\sqrt{2c}/\sigma$ [Eq.~(\ref{e.app.hopf.9})]. This means that the PDF of
$Y_t$, $f_{0}^{b} (x)$, is a symmetric extension of the ML distribution;
therefore we obtain
\begin{equation}
  f_{0}^{b} (x)
  =
  \frac {-1}{\alpha \pi}
  \sum_{k=1}^{\infty}
  \frac {\Gamma \left(\frac {k \alpha}{2} + 1 \right)}{k!}
  \left(-\frac {\sqrt{2c}}{\sigma} \right)^{k}
  |x|^{k-1} \sin \left( \frac {\pi k \alpha}{2} \right)
  \label{e.app.hopf.11}
\end{equation}
for $-\infty < x < \infty$. The PDF $f_{0}^{b} (x)$ is called the bilateral
Mittag--Leffler distribution \cite{kasahara77}. The simplest example of
this PDF is the spatial distribution of one-dimensional free CTRWs. Let the
random variable $H_k$ be the displacement of the $k$-th jump (e.g., $H_k=
\pm 1$ for CTRWs with jumps only to the nearest neighbor sites). If the
jumps are symmetric, then $\left\langle H_k \right\rangle = 0$. Thus, the
time average [Eq.~(\ref{e.app.hopf.7})] becomes a rescaled spatial position
and the PDF [Eq.~(\ref{e.app.hopf.11})] is the spatial distribution of the
free CTRWs. Note that this PDF is equivalent to the one derived from the
FFPE [Eq.~(46) in \cite{metzler00}].



%

\end {document}